\documentclass[12pt,preprint]{aastex}
\usepackage{graphicx} 
\usepackage{epstopdf}
\shorttitle{Standardizability of Type Ia Supernovae Ia in Near-Infrared}\shortauthors{Kattner et al.}

\begin{document}

\title{The Standardizability of Type Ia Supernovae in the Near-Infrared: Evidence for a Peak Luminosity Versus Decline-Rate Relation in the Near-Infrared}

\author{ShiAnne Kattner}
\affil{Department of Astronomy, San Diego State University, San Diego, CA 92182-1221, U.S.A.}
\email{skattner@sciences.sdsu.edu}

\author{Douglas C. Leonard}
\affil{Department of Astronomy, San Diego State University, San Diego, CA 92182-1221, U.S.A.}
\email{leonard@sciences.sdsu.edu}

\author{Christopher R. Burns}
\affil{Observatories of the Carnegie Institution for Science, 813 Santa Barbara St., Pasadena, CA 91101, U.S.A.}
\email{crburns@me.com}

\author{M. M. Phillips}
\affil{Carnegie Observatories, Las Campanas Observatory, Colina El Pino, Casilla 601, La Serena, Chile}
\email{mmp@lco.cl}

\author{Gast\'{o}n Folatelli}
\affil{Institute for the Physics and Mathematics of the Universe (IPMU), University of Tokyo, 5-1-5 Kashiwanoha, Kashiwa, Chiba 277-8583, Japan}
\email{gaston.folatelli@ipmu.jp}

\author{Nidia Morrell}
\affil{Carnegie Observatories, Las Campanas Observatory, Colina El Pino, Casilla 601, La Serena, Chile}
\email{nmorrell@lco.cl}

\author{Maximilian D. Stritzinger}
\affil{The Oskar Klein Centre, Department of Astronomy, Stockholm University, AlbaNova, 10691 Stockholm, Sweden}
\email{max.stritzinger@astro.su.se}
\affil{Dark Cosmology Centre, Niels Bohr Institute, University of Copenhagen, Juliane Maries Vej 30, 2100 Copenhagen \O}
\email{max@dark-cosmology.dk}
\affil{Carnegie Observatories, Las Campanas Observatory, Colina El Pino, Casilla 601, La Serena, Chile}
\email{mstritzinger@lco.cl}

\author{Mario Hamuy}
\affil{Departamento de Astronom\'{\i}a, Universidad de Chile, Casilla 36-D, Santiago, Chile}
\email{mhamuy@das.uchile.cl}

\author{Wendy L. Freedman}
\affil{Observatories of the Carnegie Institution for Science, 813 Santa Barbara St., Pasadena, CA 91101, U.S.A.}
\email{wendy@obs.carnegiescience.edu}

\author{Sven E. Persson}
\affil{Observatories of the Carnegie Institution for Science, 813 Santa Barbara St., Pasadena, CA 91101, U.S.A.}
\email{persson@obs.carnegiescience.edu}

\author{Miguel Roth}
\affil{Carnegie Observatories, Las Campanas Observatory, Colina El Pino, Casilla 601, La Serena, Chile}
\email{miguel@lco.cl}

\author{Nicholas B. Suntzeff}
\affil{George P. and Cynthia Woods Mitchell Institute for Fundamental Physics and Astronomy, Department of Physics and Astronomy, Texas A\&M University, College Station, TX 77843, USA}
\email{nsuntzeff@tamu.edu}

\begin{abstract}
We analyze the standardizability of Type Ia supernovae (SNe~Ia) in the
near-infrared (NIR) by investigating the correlation between observed
peak NIR ($YJH$) absolute magnitude and post-maximum $B$-band decline-rate
[$\Delta$$\emph{m}$$_{15}$($B$)].  A sample of 27 low-redshift SNe Ia with well-observed NIR light-curves observed by the \emph{Carnegie Supernova Project} (CSP) between 2004 and 2007 is used.  All 27
objects have pre-maximum coverage in optical bands, with a subset of 13 having
pre-maximum NIR observations as well; coverage of the other 14 begins shortly
after NIR maximum brightness.  We describe the methods used to derive light-curve
parameters (absolute peak magnitudes and decline-rates) from both spline and
template fitting procedures, and confirm prior findings that fitting templates
to SNe~Ia light-curves in the NIR is problematic due to the diversity of
post-maximum behavior of objects that are characterized by similar
$\Delta$$\emph{m}$$_{15}$($B$) values, especially at high decline-rates.  Nevertheless, we show that NIR light-curves can be reasonably fit with a template, especially if the observations begin within 5 days after NIR maximum. 
SNe~Ia appear to be better ``standardizable candles"  in the NIR bands than in the optical bands.  For the subset of 13 objects in our data set that excludes the highly reddened and fast-declining SNe~Ia, and includes only those objects for which NIR observations began prior to 5 days after maximum light, we find modest (1.7$\sigma$) evidence for a peak luminosity vs. decline-rate relation in $Y$, and stronger evidence (2.8$\sigma$) in $J$ and $H$.  Using $\emph{R$_V$}$
values differing from the canonical value ($\emph{R$_V$}$ = 3.1) is shown to have little
effect on the results.  A Hubble diagram is presented for the NIR bands and the $B$-band.  The resulting scatter for the combined NIR bands is 0.13 mag, while the $B$-band produces a scatter of 0.22 mag.  Finally, we find evidence for a bimodal distribution in the NIR
absolute magnitudes of fast-declining SNe~Ia [$\Delta$$\emph{m}$$_{15}$($B$) $>$
1.7].  These data suggest that applying a correction to SNe~Ia peak luminosities
for decline-rate is likely to be beneficial in the $J$- and $H$-bands to make SNe~Ia more
precise distance indicators, but of only marginal importance in the $Y$-band.
\end{abstract}


\section{Introduction}
It is widely accepted that SNe~Ia are excellent $\emph{standardizable candles}$
at optical wavelengths.  After applying an empirical correction established
between light-curve shape and peak magnitude, SNe~Ia become one of the most
precise extragalactic distance indicators known.  \cite{1993ApJ...413L.105P} was the first to
discover a tight correlation (hereafter the ``Phillips method'') between
optical absolute magnitudes at maximum light and the decline-rate parameter
$\Delta$$\emph{m}$$_{15}$($B$), defined as the drop in $B$-band brightness from peak to
15 days later.  Slow-declining SNe Ia [i.e., lower $\Delta$$\emph{m}$$_{15}$($B$)] are intrinsically brighter than their fast-declining counterparts.  Since Phillips' initial work, the Phillips method has continued to be
utilized \citep{1996AJ....112.2391H,1999AJ....118.1766P,2006ApJ...647..501P,2011AJ....141...19B},
along with other similar calibration methods, such as the ``stretch" method
\citep{1997ApJ...483..565P,2001ApJ...558..359G}, the multicolor light-curve
shape method (MLCS/MLCS2k2; \citealt{1996ApJ...473...88R,1998AJ....116.1009R};
  \citealt{2007ApJ...659..122J}), the color-magnitude intercept calibration method
(CMAGIC; \citealt{2003ApJ...590..944W}), the spectral adaptive light-curve template method (SALT/SALT2; \citealt{2005A&A...443..781G,2007A&A...466...11G}), and SiFTO (\citealt{2008ApJ...681..482C}).  Applying such corrections decreases
the dispersion in measured distance estimates to the 0.20 mag level or less (\citealt{1996AJ....112.2391H, 2006ApJ...647..501P, 2008ApJ...689..377W, 2010AJ....139..120F}), which
significantly improves the accuracy of SNe~Ia as standardizable candles at
optical wavelengths.

SNe~Ia have played a critical role in cosmological studies over the past two
  decades.  While the majority of SNe~Ia studies have been performed using
  optical bands, there has long been the hope that they might actually have greater accuracy in NIR bands, due to the reduced effects of dust extinction, which is one of the main sources of error in distance determinations (\citealt{2009ApJ...704.1036F}).  This has motivated recent work on establishing the
  standardizability of SNe~Ia in the NIR, which has been shown to be advantageous for several reasons.  Extinction corrections are smaller, on the order of a magnitude less, in NIR than in optical bands (\citealt{1989ApJ...345..245C}), and there appears to be a shallower dependence of absolute NIR magnitude on decline-rate compared with optical bands (Krisciunas et al. 2004a,c; \citealt{2008ApJ...689..377W}; \citealt{2010arXiv1011.5910M}; \citealt{2010AJ....139..120F}).  Recently, \cite{2008ApJ...685..752G}, \cite{2010MNRAS.406..782S}, and \cite{2010ApJ...715..743K} have found that the absolute magnitudes of SNe~Ia depend on host-galaxy properties (i.e., star formation rate, host-galaxy mass, host-galaxy metallicity), making the correction factors evolve with redshift.  Therefore, minimizing these correction factors by observing in the NIR is important, not only to decrease the random errors incurred for each SN~Ia's correction, but also to reduce our sensitivity to any systematic evolution of the correction factors with redshift.
  
The first NIR observation of a SN~Ia was reported by
  \cite{1973ApJ...180L..97K}, and the first extensive NIR data set was
  obtained by Elias et al. (1981, 1985).  Since that
  time, there has been a large increase in the number of SNe~Ia light-curves
  observed at NIR wavelengths, especially within the last decade (e.g.,
  \citealt{1987ApJ...315L.129F}; \citealt{1999ApJS..125...73J}; \citealt{2003fthp.conf..193P}; Krisciunas et
  al. 2004a,b,c, 2006; \citealt{2007A&A...470L...1S}; \citealt{2008ApJ...689..377W};
  \citealt{2010AJ....139..519C}; \citealt{2011AJ....142..156S} and references therein).  Two groups, in particular, have been working to obtain homogenous optical and NIR data sets of SNe of all types in multiple filters: the CfA Supernova Group (\citealt{2008ApJ...689..377W}) and the \emph{Carnegie Supernova Project} (CSP; \citealt{2006PASP..118....2H}).
  
Initial studies suggested that SNe~Ia may present a weaker peak
  luminosity vs. decline-rate dependence in the NIR compared with the optical,
  which display slopes of 0.63, 0.61, 0.57, and 0.52 mag unit$^{-1}$ decline-rate [as measured by $\Delta$$\emph{m}$$_{15}$($B$)] in $BVRI$,
  respectively (\citealt{2006ApJ...647..501P}).  
This led some early investigators (e.g.,
  \citealt{1985ApJ...296..379E,1993ApJ...413L.105P,2000MNRAS.314..782M}) to
  suggest that these objects may be nearly perfect standard candles in the NIR.
  Krisciunas et al. (2004a,c) confirmed these suggestions, finding no obvious
  decline-rate relations in the $JHK_s$ bands, concluding that SNe~Ia might
  well be nearly perfect standard candles in the NIR at the $\pm$ 0.20 mag level or better.  
A study by \cite{2008ApJ...689..377W}, which did not correct for optical light-curve shape, found SNe~Ia to be excellent
  standard candles in the NIR, deriving an intrinsic dispersion in absolute
  magnitude of only 0.28 mag in the $J$-band and 0.15 mag in the $H$-band.  \cite{2010AJ....139..120F} found a marginal dependence of absolute NIR magnitudes on decline-rate using CSP SNe Ia (0.44 $\pm$ 0.14, 0.58 $\pm$ 0.09, 0.33 $\pm$ 0.18 mag unit$^{-1}$ decline-rate in $YJH$, respectively).
  Some of these earlier studies, however, employ inhomogeneous
  samples of SNe~Ia (e.g., different telescopes, different reduction
  procedures), so subtle systematics may complicate interpretation of the results.
  
In this article we re-examine the standardizability of SNe~Ia in the NIR bands using a
  homogeneously obtained sample of 27 low-redshift objects observed by the CSP between
  2004 and 2007.  We use these data to quantitatively examine whether a decline-rate dependence correction is needed in the NIR or whether
  SNe~Ia are indeed perfect standard candles at these wavelengths.  A detailed
  description of the CSP observations, data reduction, and photometry processes
  can be found in \cite{2006PASP..118....2H} and \cite{2010AJ....139..519C}.
  All of the photometric data are on the Swope+CSP natural photometric system,
  with the final SN photometry published by \cite{2010AJ....139..519C}, with
  the exception of three SNe~Ia (SNe~2006et, ~2007af, and ~2007on) whose
  photometry are presented in the second CSP data release article
  (\citealt{2011AJ....142..156S}).
  
This article is organized as follows. In \textsection{2} we briefly discuss
  light-curve template-fitting using SNooPy (SuperNova in Object-Oriented
  Python; \citealt{2011AJ....141...19B}), and we point out difficulties with the
  template-fitting approach.  In \textsection{3} we examine the standardizability of SNe~Ia in NIR bands by examining the efficacy of the
  Phillips method to fit the relationship between absolute peak $Y, J,
  {\rm\ and\ } H$ magnitudes and decline-rates.  In \textsection{4} we investigate the bimodal distribution of the fast-declining, low-luminosity SNe at these wavelengths.  We summarize our conclusions in
  \textsection{5}.

\section{NIR Light-Curve Fitting and Morphology}
\subsection{Light-Curve Fitting Using SNooPy}
From 2004--2007, the CSP obtained $\sim 100$ SN~Ia light-curves in both optical
and NIR bands.  Here, we choose 27 of the best-observed objects discovered during the first three campaigns, all of which have
optical observations that start before maximum light and continue for at least
30 days post-maximum.  Thirteen of the SNe~Ia have NIR observations that begin
before maximum brightness as well, while the others begin shortly after maximum
brightness.  We denote a best-fit (BF) subsample as those 13
objects with observed maxima in both optical and NIR bands.  This BF subsample is used
to create a ``training set'' for the template light-curves following the
process described here.

SN~Ia light-curve parameters are typically measured with respect to the peak of
the light-curve.  We prefer to directly measure the $\emph{K}$-corrected, de-reddened
time of maximum ($\emph{t}$$^{max}_{X}$), peak magnitude ($\emph{m}$$^{max}_{X}$), and
decline-rate [$\Delta$$\emph{m}$$_{15}$($B$)] based on a cubic-spline interpolation of the data points themselves (see the discussion in \citealt{2011AJ....141...19B}).  However, in $\emph{YJHK$_s$}$,
SNe~Ia typically achieve peak brightness 3--4 days prior to $B$-band maximum,
which often results in having objects with well-observed optical peaks but NIR
observations that begin post-peak.  Although not optimal, peak values (i.e., $\emph{t}$$^{max}_{X}$ and $\emph{m}$$^{max}_{X}$) can be estimated for such objects by
fitting their post-peak data with a template light-curve based on observations
of SNe~Ia with well-observed optical and NIR peaks.

To generate and fit our templates, we employ the SNooPy package
(\citealt{2011AJ....141...19B}).  To create the training set using the best-observed objects, SNooPy examines the $B$-band light-curves and computes
where the derivative is zero on the cubic-spline fit.  This point is used to
estimate $\emph{t}$$^{max}_{X}$ and $\emph{m}$$^{max}_{X}$ (and their uncertainties).
Once $\emph{t}$$^{max}_{B}$ is found, SNooPy measures $\Delta$$\emph{m}$$_{15}$($B$) and then
$\emph{m}$$^{max}$ in all of the other filters.  This training set of SNe~Ia is then used
to create the template $ugriBVYJH$ light-curves through the procedure detailed by
\cite{2011AJ....141...19B}.  Briefly, $\emph{t}$$^{max}$, $\emph{m}$$^{max}$, and
$\Delta$$\emph{m}$$_{15}$($B$) of the training set are placed on a three-dimensional
surface for each band.  SNooPy interpolates across the surface along a constant
$\Delta$$\emph{m}$$_{15}$($B$) line using a 2D variation of the Gloess algorithm
(\citealt{2004AJ....128.2239P}).  From the template light-curves, SNooPy then
measures the best-fit values of $\emph{t}$$^{max}$, $\emph{m}$$^{max}$, and $\Delta$$\emph{m}$$_{15}$, which is the template-derived value of the decline-rate derived from a combination of all filters,
via $\chi$$^2$ minimization.  For consistency, we use $\Delta$$\emph{m}$$_{15}$($B$), which is the decline-rate directly measured from the $B$-band, as the decline-rate parameter for all of our objects in all filters, including those objects that were template fit.  The measured $\Delta$$\emph{m}$$_{15}$($B$) parameters for all SNe are listed in Table~\ref{tab:tab1}.

\subsection{Light-Curve Morphology and the Second Maximum in the NIR}
\begin{figure}
\begin{center}
\scalebox{0.75}{
\plotone{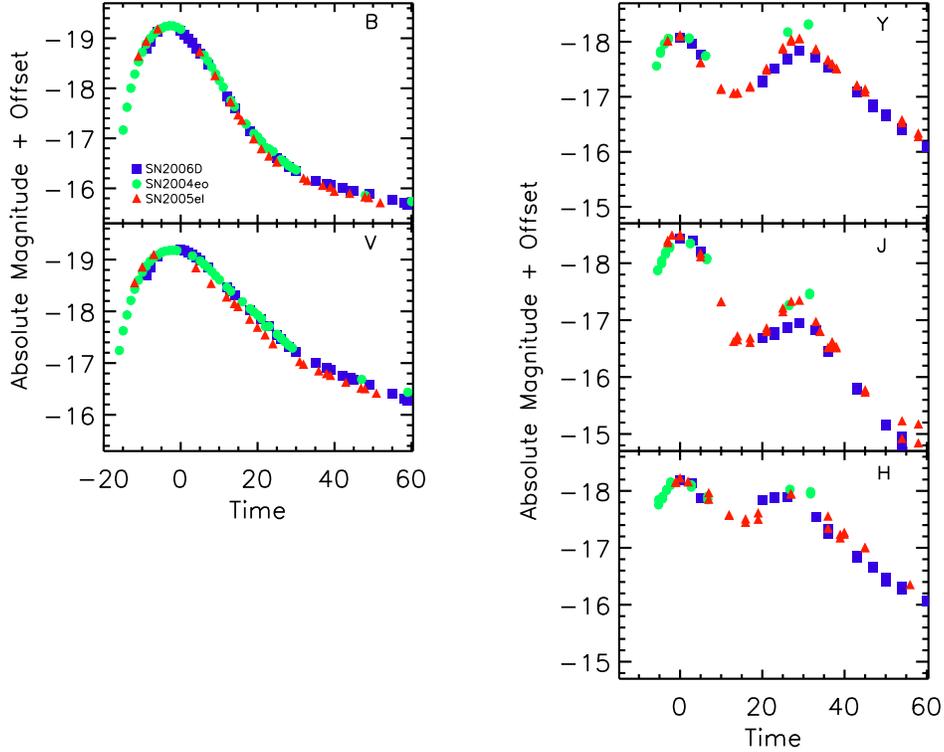}
	}
\end{center}
\caption{\emph{Left}: Absolute magnitude $B$ and $V$ light-curves of SNe~2006D, 2004eo, and  2005el.  The light-curves are shifted such that maxima agree.  All SNe have similar decline-rates in the range of $\Delta$$\it{m}$$_{15}$($B$)=1.35 $-$ 1.39 mag.  Blue squares represent SN~2006D, green circles represent SN~2004eo, and red triangles represent SN~2005el.  Note the similarity in the $B$ and $V$ light-curves.  \emph{Right}: Absolute magnitude $YJH$ light-curves of the same SNe.  Again, the light-curves are shifted to a common maximum.  Note the difference in the $YJH$ light-curves, especially the difference in strength of the second maxima in the $J$-band.
\label{fig:fig1} }
\end{figure}

\begin{figure}
\begin{center}
\scalebox{0.75}{
\plotone{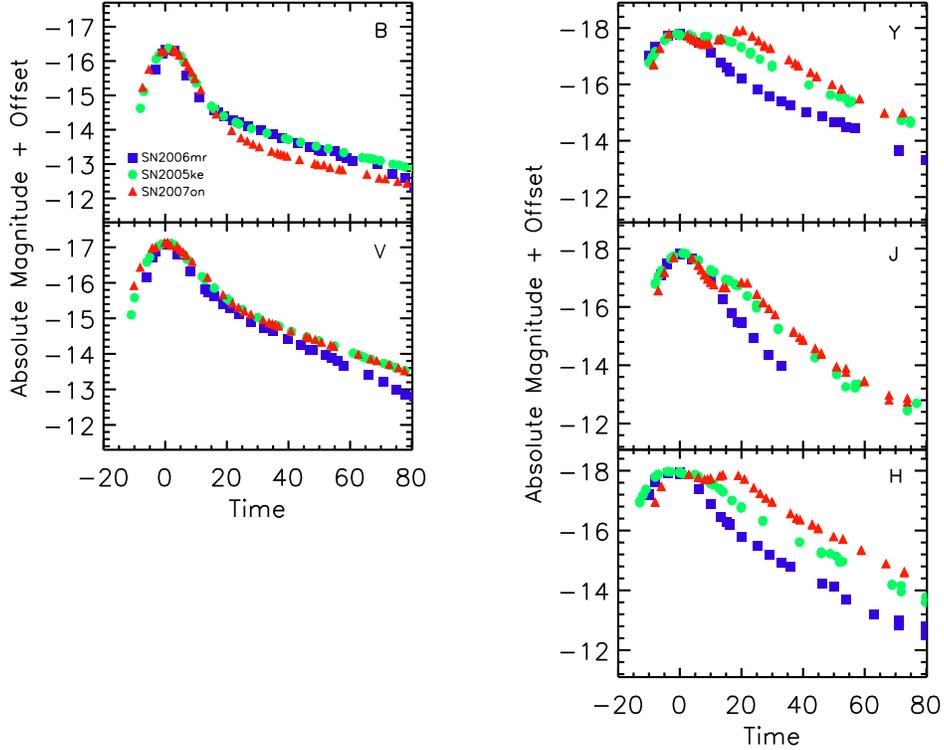}
	}
\end{center}
\caption{\emph{Left}: Absolute magnitude $B$ and $V$ light-curves of SNe~2006mr, 2005ke, and 2007on ($\Delta$$\it{m}$$_{15}$($B$)=1.75-1.89).  Blue squares represent SN~2006mr, green circles represent SN~2005ke, and red triangles represent SN~2007on.  The light-curves are shifted to a common maximum.  As in Fig. \ref{fig:fig1}, the optical light-curves are very similar.  \emph{Right}: Absolute $YJH$ magnitude light-curves of the same SNe.  The light-curves are shifted to a common magnitude.  Note the large difference in the NIR light-curves, especially between SN~2007on, which shows a rise to second maxima.
\label{fig:fig2} }
\end{figure}

Unlike optical light-curves, which are quite homogeneous for a fixed
$\Delta$$\emph{m}$$_{15}$($B$), light-curves at longer wavelengths display a more
diverse morphology, particularly due to the existence of a rise to a ``second
maximum'' following the initial peak.  Investigation in the $\it{i}$-band has shown that this second peak can exhibit a strength and
morphology that varies significantly from SN~Ia to SN~Ia with identical
$\Delta$$\emph{m}$$_{15}$($B$) values (e.g., \citealt{2009ApJ...704.1036F}; \citealt{2010AJ....139..120F}; \citealt{2011AJ....141...19B}).  We confirm this with our sample in $YJH$ (see Figs.
\ref{fig:fig1} and \ref{fig:fig2}).  Note the similarity in the $B$ and $V$ light-curves in Figures \ref{fig:fig1} and \ref{fig:fig2} and the difference in  the $YJH$ light-curves, especially around the second maximum. This creates a problem for template fitting NIR
light-curves using a one-parameter descriptor of
light-curve shape, especially for fast-declining SNe Ia [i.e., $\Delta$$\emph{m}$$_{15}$($B$) $>$ 1.7].

\begin{figure}
\begin{center}
\scalebox{0.75}{
\plotone{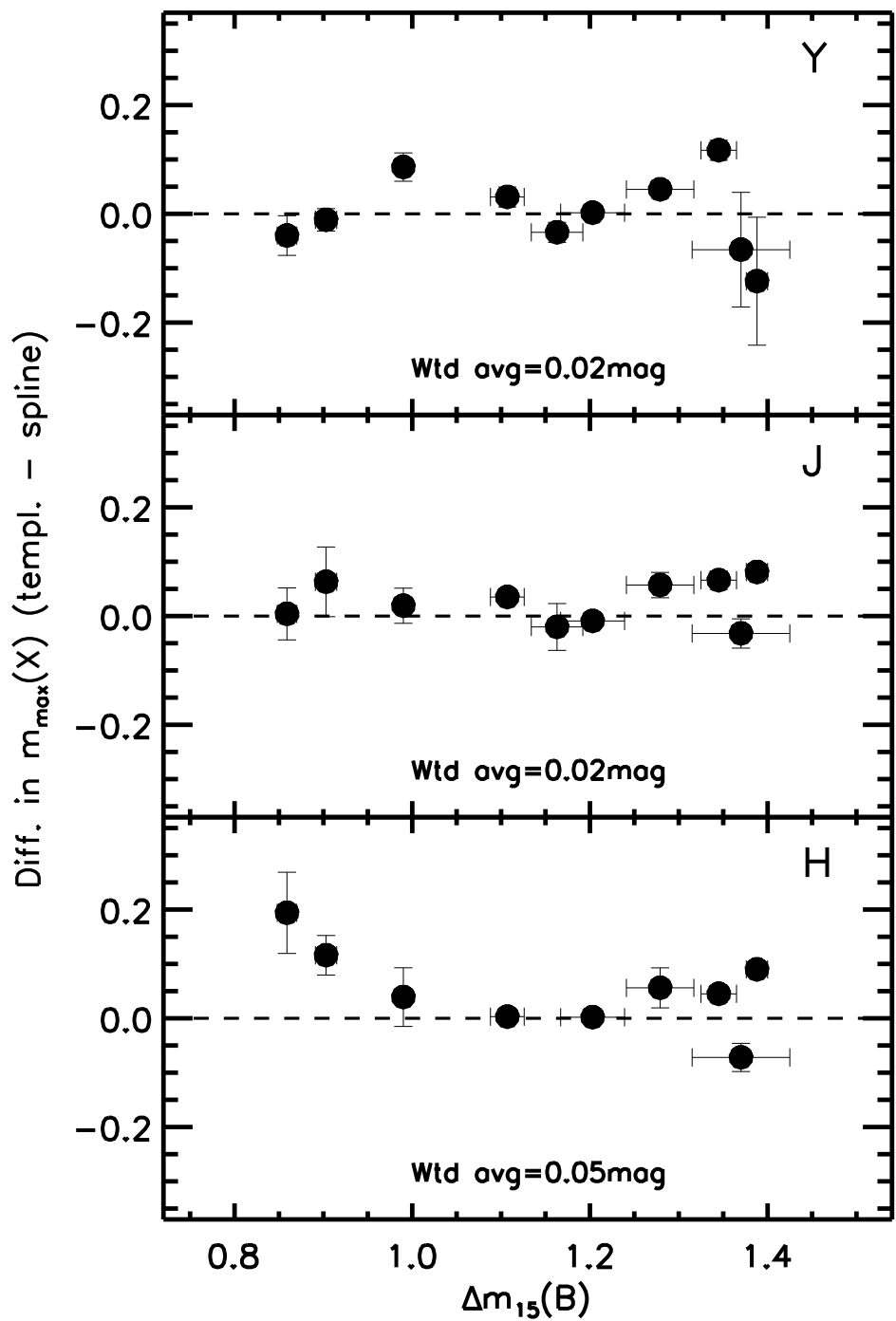}
	}
\end{center}
  \caption{Plot of the difference in peak magnitude derived from template vs. spline fits in $YJH$ as a function of decline-rate for objects with observed NIR maxima.  Uncertainties are smaller than the points, unless shown.  Evidence of systematic differences are present, but the systematic errors are not large, compared with the uncertainties in the final absolute magnitudes.  
\label{fig:fig3} }
\end{figure}

To test the accuracy of light-curve parameters derived from template fits, we compare the peak $YJH$
magnitudes obtained from template fits with those obtained directly from spline
fits for the BF group [excluding SNe with $\Delta$$\emph{m}$$_{15}$($B$) $>$ 1.7] in Figure~\ref{fig:fig3}.  The weighted averages of the difference in peak magnitude for $YJH$ are 0.02 mag, 0.02 mag, and 0.05 mag, respectively, for objects with $\Delta$$\emph{m}$$_{15}$($B$) $<$ 1.7.  Figure
\ref{fig:fig3} shows evidence of systematic differences, particularly in $H$, but the systematic errors are not large compared with the uncertainties in the final absolute magnitudes due to the errors in the host color excess and the derived distances.  

\cite{2010AJ....139..120F} also found evidence of systematic differences in peak magnitudes derived from template vs. spline fits for the NIR filters, as shown in their Figure 5, stating that the poor precision of the template fit in $iYJH$ is partly due to the small sample used to derive the templates and also due to the variation in morphology surrounding the second maximum.
\cite{2011AJ....141...19B} used a bootstrap technique to incorporate the extra dispersion found in the template fits caused by NIR light-curve variations.  We have attempted to account for this extra dispersion by adding the extrapolation errors derived by \cite{2011AJ....141...19B} in quadrature with the uncertainties in the apparent magnitude obtained from the SNooPy fits.

\begin{figure}
\begin{center}
\scalebox{0.75}{
\plotone{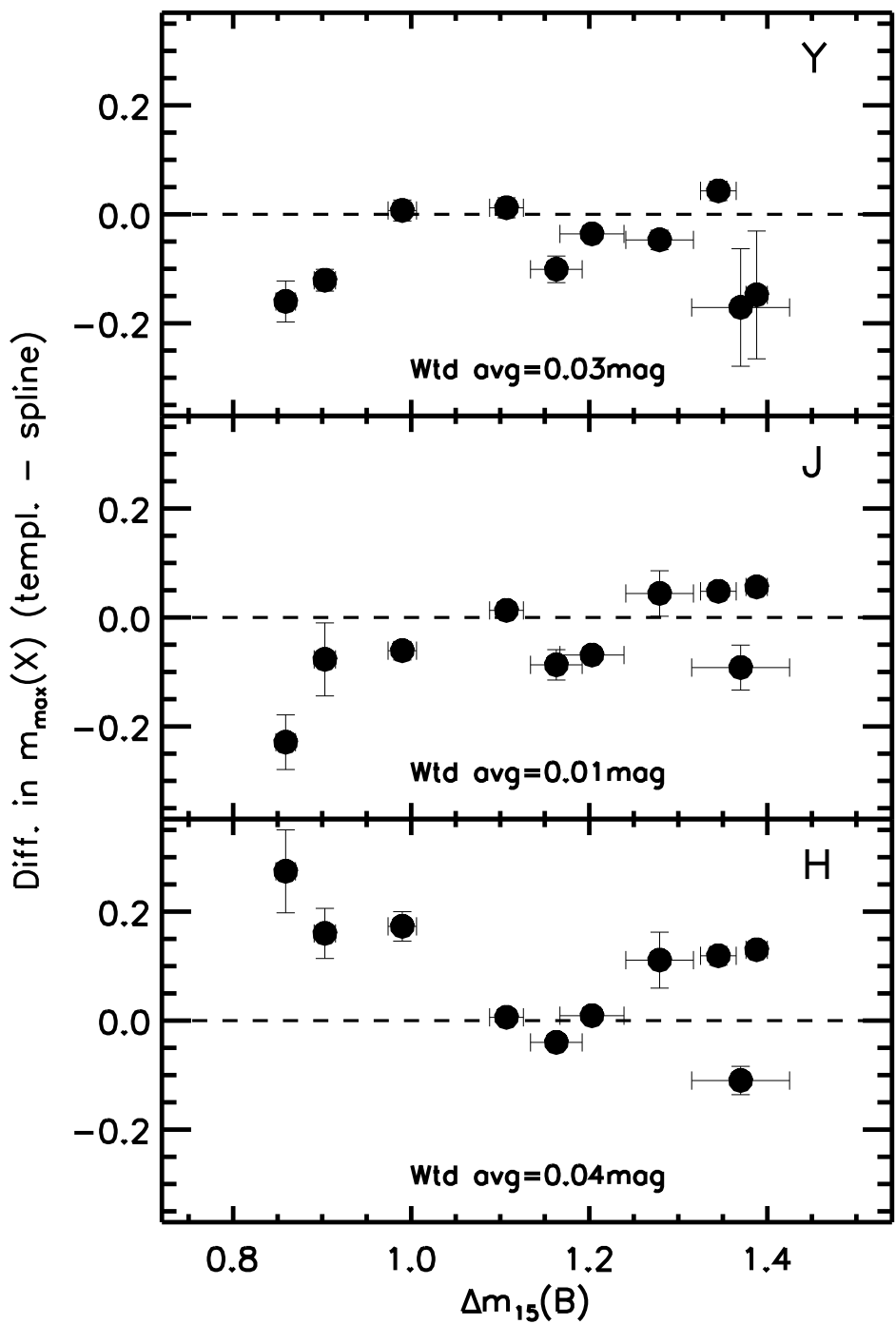}
	}
\end{center}
  \caption{Plot of the difference in peak magnitude derived from template vs. spline fits in $YJH$ as a function of decline-rate for SNe whose NIR observations begin within 5 days after NIR maximum.  Uncertainties are smaller than the points, unless shown.  The weighed averages presented here are about the same as the weighted averages found in Fig. \ref{fig:fig3}, suggesting that when the observations begin within 5 days of NIR maximum, SNooPy does an adequate job of deriving peak light-curve parameters from template fits.  
\label{fig:fig4} }
\end{figure}

Because the variations in the strength of the secondary maximum affect the accuracy of light-curve      template fitting in the $iYJH$ bands, we might expect the uncertainties in the peak NIR magnitudes to be a function of how many days past maximum the observations begin.  \cite{2010AJ....139..120F} found that if a SNe~Ia has observations that begin within $\sim$1 week after the time of maximum, the random uncertainty in peak magnitude is $\sim$0.1 mag and the systematic uncertainty is only $\sim$0.03 mag; a template fit for a SNe~Ia with photometry that starts later than this will be unreliable.  Two thirds of the events in our sample with $\Delta$$\emph{m}$$_{15}$($B$) $<$ 1.7 have photometry that starts within 5 days of the NIR maximum, and so it is interesting to see if the template fitting procedure gives reasonable estimates of the peak magnitude for these SNe~Ia.  To test this, we create a plot similar to that of Figure \ref{fig:fig3}.  For this test, we re-derive the template fit peak NIR apparent magnitudes by removing all of the data prior to 5 days after NIR maximum for each SN~Ia and running the data through SNooPy again.  Figure \ref{fig:fig4} shows the resulting plots in $YJH$.  There appear to be some systematic differences, but, again, they are not large when compared wih the uncertainties in the final absolute magnitudes.  The weighted averages of the differences are 0.03 mag in $Y$, 0.01 mag in $J$, and 0.04 mag in $H$, which is about the same for the weighted averages found in Figure \ref{fig:fig3}, suggesting that when the observations begin within 5 days of NIR maximum, SNooPy does an adequate job of deriving peak light-curve parameters from template fits.  

It is unclear at this point how to best handle the diversity of light-curve
morphologies in the NIR when applying templates to SNe~Ia whose NIR observations start more than 5 days after NIR maxima.  Introducing a second
parameter, such as another peak magnitude, decline-rate relation (such as a $\Delta$m$_{15}$-like parameter defined for the $YJH$ bandpasses), host-galaxy property, or spectral feature, might help (e.g., \citealt{2006ApJ...649..939K}; \citealt{2009ApJ...699L.139W}; \citealt{2011ApJ...729...55F}; \citealt{2010MNRAS.406..782S}; \citealt{2011A&A...526A..81B}), but identifying this second parameter is difficult with the limited number of well-observed events.  With significant differences among the NIR light-curves of SNe~Ia with
similar $\Delta$$\emph{m}$$_{15}$($B$) values, though, template fitting in $iYJH$ based on
the $\Delta$$\emph{m}$$_{15}$($B$) parameter alone may be subject to significant
uncertainties.  Here, we proceed
to cautiously apply templates when needed, while using the BF subsample of objects, for which no template-fitting is
needed, to check our results. 

\section{Investigating the Homogeneity of Peak Luminosity of SNe~Ia in the NIR}
\subsection{Distance Moduli, Color Excess, and Reddening}
In this section we use our homogenous sample of light-curves to examine the
precision of SNe~Ia as standard candles in the NIR bands.  We calibrate the
absolute magnitudes of our objects using a technique first proposed by
\cite{1999AJ....118.1766P}, which examines the correlation between
reddening-corrected, absolute peak magnitudes vs. decline-rate according to
the model:
\begin{equation}\mu_X = m_X - M_X(0) - b_X[\Delta m_{15}(B) - 1.1] - R_{X} E(B-V),\end{equation}\\
where $\mu$$_{X}$ is the distance modulus, $\emph{m}$$_X$ is the peak apparent magnitude
corrected for Galactic reddening given in band $X$, $\emph{M}$$_X$(0) is the peak absolute
magnitude for $\Delta$$\emph{m}$$_{15}$($B$)=1.1 and zero color excess, $\emph{b}$$_X$ is the slope of the
luminosity vs. decline-rate relation, $\emph{R}$$_X$ is the total-to-selective absorption
coefficient given in band $X$, and $\emph{E(B-V)}$ is the SN color excess due to host-galaxy dust.  

To calculate the distance modulus, which is used to transform rest-frame
apparent magnitude to absolute magnitude, we employ the following
approximation to the luminosity distance:
\begin{equation}d_{L}(z_{CMB}; H_0, \Omega_M, \Omega_\Lambda) =\frac{(1 + z_{helio})}{(1 + z_{CMB})} \frac{c}{H_0} [z_{CMB} + \frac{1}{2}(\Omega_\Lambda - \frac{\Omega_M}{2} + 1) z_{CMB}^2],\end{equation}
where $\emph{z}$$_{helio}$ is the heliocentric redshift of the host-galaxy; $\emph{z}$$_{CMB}$
is the redshift of the host-galaxy in the cosmic microwave background (CMB) rest frame; and the standard
cosmological parameters are $\emph{H}$$_0$ = 72 km s$^{-1}$ Mpc$^{-1}$
(\citealt{2001ApJ...553...47F}), $\Omega$$_M$ = 0.28, and $\Omega$$_{\Lambda}$
= 0.72 (\citealt{2007ApJS..170..377S}).  Values of $\emph{z}$$_{helio}$ and $\emph{z}$$_{CMB}$
are given in Table~\ref{tab:tab1}.  The uncertainty in the velocity of the
host-galaxies due to peculiar velocity is assumed to be $\sigma$$_z$=0.001 (300
km s$^{-1}$).  

For SNe that are not in the smooth Hubble flow (i.e., $z \lesssim 0.01$), distances derived
using equation (2) can be inaccurate.  Seven of our objects fall into this
category: SNe~2005am, ~2005ke, ~2006D, ~2006X, ~2006mr, ~2007af, and
~2007on.  Four of these (SNe~2005ke, ~2006X, ~2006mr, and ~2007on) have
published Cepheid or surface brightness fluctuation distances for their host-galaxies, which were used (see Table~\ref{tab:tab2}); the other three objects were
excluded from our sample, leaving a sample of 24 SNe~Ia.

For our complete sample, we employ a mix of cubic-spline (when peak is observed) and
template (when necessary) fits to the light-curves to derive the parameters of $\emph{m}$$^{max}$ and $\Delta$$\emph{m}$$_{15}$($B$).  Table~\ref{tab:tab1} lists the fitting method for each SN~Ia, and Table~\ref{tab:tab3} lists the $\emph{K}$-corrected spline and template-derived apparent peak magnitude in $BVYJH$ for each object.  $\emph{K}$-corrections are applied to convert the observed magnitudes to rest-frame magnitudes using the \cite{2007ApJ...663.1187H} spectral templates.  For further details on the $\emph{K}$-correction procedure, see \cite{2011AJ....141...19B}.

All peak magnitudes have been corrected for Galactic reddening using the values
${E(B-V)}$$_{gal}$ given in Table~\ref{tab:tab1} (\citealt{1998ApJ...500..525S}) and adopting $\emph{R}$$^{Gal}_V$=3.1
(a further discussion on this point follows).  In order to estimate extinction in the host-galaxies, we have used color excesses $E(B-V)$ obtained from the observed SNe colors as follows.  For SNe with
$\Delta$$\emph{m}$$_{15}$($B$) $<$ 1.7, the host-galaxy color excesses in column (7) of
Table~\ref{tab:tab1} are derived using the intrinsic $\emph{B}$$_{max}$--$\emph{V}$$_{max}$ color law at maximum light derived by
\cite{2010AJ....139..120F}.  
The Folatelli et al. relation is only valid for $\Delta$$\emph{m}$$_{15}$($B$) $<$ 1.7; therefore, to calculate $\emph{E(B-V)}$$_{host}$ for the three fast-declining events, the re-derived CSP Lira law (1995) described by \cite{2010AJ....139..120F} is used.  We do not apply any priors to the color excess measurements; therefore, negative reddenings are possible and, indeed, expected due to the measurement uncertainty and intrinsic color dispersion in SNe~Ia. 

Significant debate exists on the ``best'' value of the total-to-selective
absorption coefficient for the host-galaxy, $\emph{R}$$_V$, for SNe~Ia studies.  By reducing the scatter in
the Hubble diagram, some studies (e.g.,
\citealt{1999ApJ...525..209T,2006ApJ...645..488W,2006A&A...447...31A,2006MNRAS.369.1880E,2007ApJ...664L..13C,2008A&A...487...19N})
find that SNe~Ia ``prefer'' a lower $\emph{R}$$_V$ value ($\emph{R}$$_V$=1.0-2.4) compared with the value obtained through typical lines of sight in the Milky Way ($\emph{R}$$_V$=3.1).  Other studies (\citealt{2009ApJ...699L.139W}; \citealt{2010AJ....139..120F};
\citealt{2011ApJ...729...55F}; \citealt{2010arXiv1011.5910M}) that examine the colors of SNe~Ia as a function of wavelength find that the choice
between a high or low $\emph{R}$$_V$ value depends on the reddening of the particular
SN.  It appears that SNe~Ia that suffer from minimal$-$to$-$moderate extinction tend to favor a value close to $\emph{R}$$_V$=3.1, while those that suffer from high extinction favor a lower value.
These studies
conclude that the low derived $\emph{R}$$_V$ values are, in all likelihood, {\it not}
physically due to changes in the actual reddening coefficient, but rather to
some as yet-unknown parameter affecting the intrinsic color of SN~Ia light-curves (see, however, \citealt{2009ApJ...699L.139W}).  

The reason for these differing $\it{R}$$_V$ values is not well understood, and the exact value depends on the method used to derive it.  Given our specific goal of removing the effects of dust extinction from SN~Ia
light-curves in the most accurate (i.e., physically understood) manner
possible, we choose to use the typical Galactic value of $\it{R}$$_V$=3.1.  We
emphasize that our primary purpose in this study is to examine the dependence
of intrinsic (i.e., corrected only for extrinsic effects, such as dust
extinction) peak absolute magnitude on decline-rate in the NIR bands -- not
to minimize dispersion in the Hubble diagram or to calculate cosmological
parameters; thus, choosing $\it{R}$$_V$=3.1 is warranted.  As we shall see, since
the effects of dust extinction in the NIR are minimal, this choice is not
crucial for the results obtained.  Indeed, as pointed out by \cite{2000ApJ...539..658K} and \cite{2010AJ....139..120F}, this is one of the significant advantages of working in the NIR.

\subsection{Luminosity Versus Decline-Rate Relation}
In this section we investigate the correlation between peak luminosity and
decline-rate for 5 different subsamples (see Table~\ref{tab:tab4}).  We investigate the 5 subsamples in order to isolate the possible systematic effects that individual SNe Ia may have on the fit.  Since each $\it{R}$$_X$ (being a function only of $\it{R}$$_{V}$) is fixed, we solve only for $\it{M}$$_X$(0) and $\it{b}$$_X$ using
equation (1) in the following analyses. \\

\begin{figure}
\begin{center}
\scalebox{0.75}{
\plotone{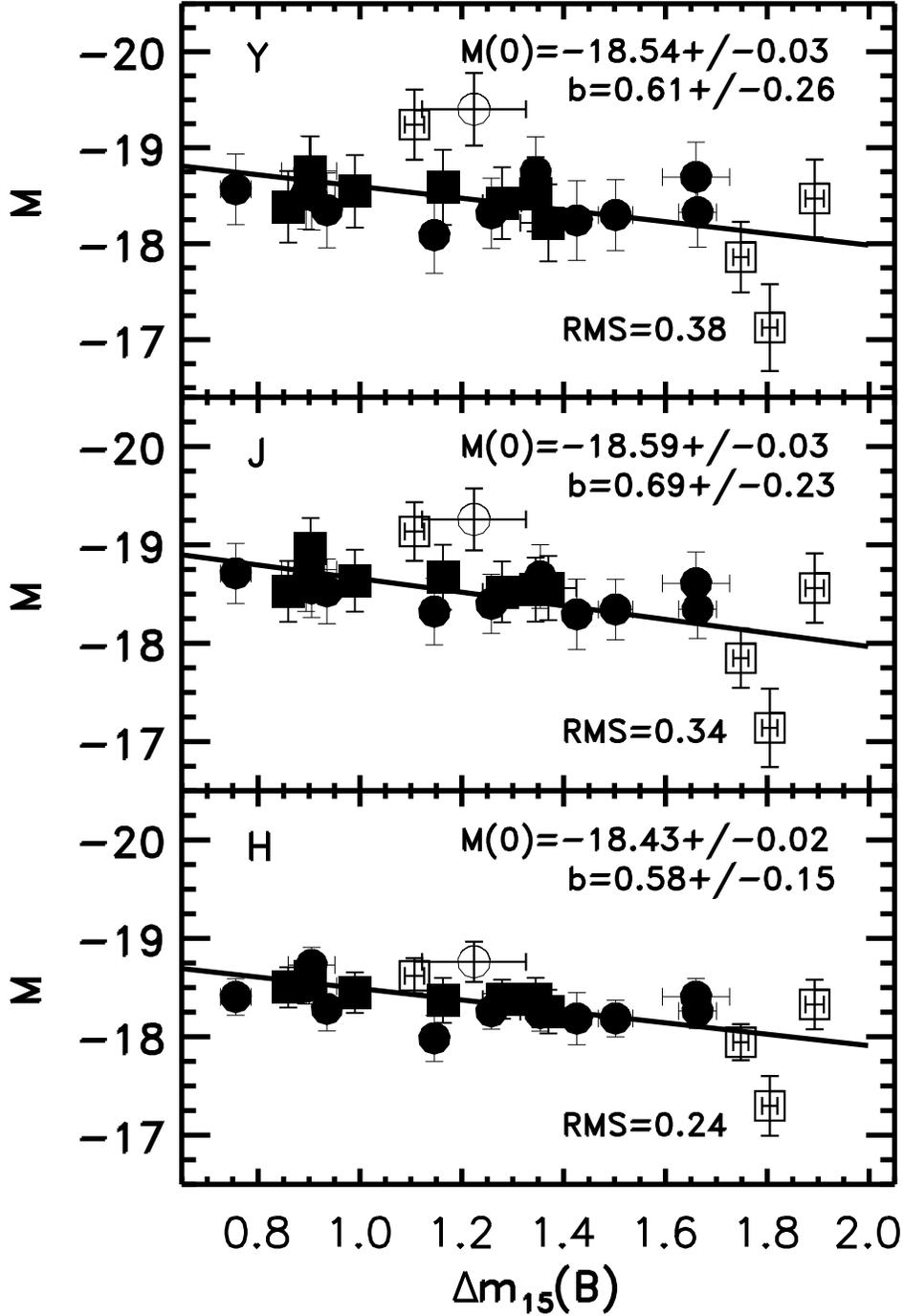}
	}
\end{center}
\caption{Fits of absolute $YJH$ magnitude vs. decline-rate for subsample 1.  Filled circles represent the SNe fit with a template, filled squares represent SNe fit with a spline.  Open symbols represent fast-declining SNe and the highly reddened SNe.  Uncertainties associated with absolute magnitude are smaller than the points, unless shown.  All three of the NIR bands show a statistically
significant luminosity vs. decline-rate correlation, with a dependence in
all three NIR bands comparable with those found in optical bands.
\label{fig:fig5} }
\end{figure}

{\it Subsample 1}.  Includes all 24 SNe~Ia (i.e., the complete sample of 27 SNe Ia minus the three for which distances are not available [see $\S$3.1]).  As shown in Table~\ref{tab:tab5} and displayed in
Figure~\ref{fig:fig5}, all three of the NIR bands show a statistically
significant luminosity vs. decline-rate correlation, with a dependence in
all three NIR bands comparable with those in the $B$-band found by Prieto et
al. (2006).  However, as seen from the next subsample, the
fast-declining and highly reddened objects have a very large effect on the
derived slope.

\begin{figure}
\begin{center}
\scalebox{0.75}{
\plotone{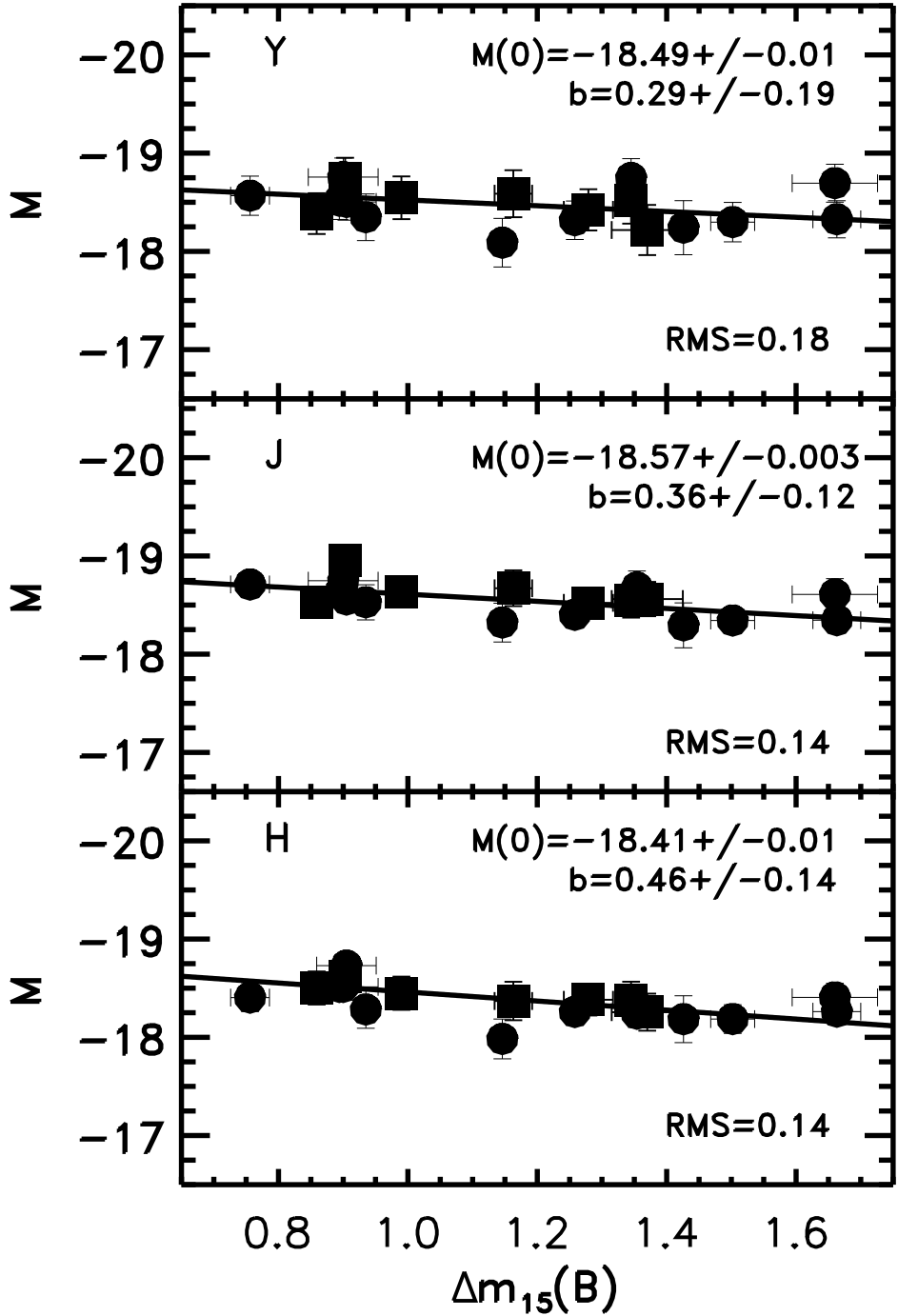}
	}
\end{center}
\caption{Fits of absolute $YJH$ magnitude vs. decline-rate for subsample 2.  The symbols for
  points are the same as those used in Fig. 5.  A significantly weaker dependence of absolute peak magnitude on decline-rate is found for the $YJH$ bands for subsample 2, and the rms dispersion decreasing significantly in all bands as well.
\label{fig:fig6} }
\end{figure}

{\it Subsample 2}.  Excludes the three fast-declining SNe~Ia and the two most highly reddened objects (SNe 2005A and 2006X).  The highly reddened objects were excluded in order to minimize the systematic uncertainties due to the apparent peculiar nature of the reddening that affects them (\citealt{2010AJ....139..120F}).  We choose to remove the SNe~Ia with $\Delta$$\it{m}$$_{15}$($B$) $>$ 1.7 because this decline-rate cutoff is widely used in other peak luminosity vs. decline-rate studies (Hamuy et al. 1996a; \citealt{1999AJ....118.1766P}; Krisciunas 2004a,c; \citealt{2006ApJ...647..501P}; \citealt{2010AJ....139..120F}), and as found by \cite{2009AJ....138.1584K}, such objects appear to obey a different peak luminosity vs. decline-rate relation compared with normal SNe~Ia and should therefore be modeled separately. 
Omitting the highly reddened and fast-declining events creates a shallower slope on the luminosity vs. decline-rate relation compared with subsample 1 for all three NIR bands
(Fig.~\ref{fig:fig6}), with the rms dispersion decreasing
significantly in all bands as well.  There is evidence for a weak correlation between peak luminosity and decline-rate in $Y$, with the significance of the measured slope being 1.5$\sigma$, and a strong correlation in both $J$ and $H$ at the 3$\sigma$ significance level. 

\begin{figure}
\begin{center}
\scalebox{0.75}{
\plotone{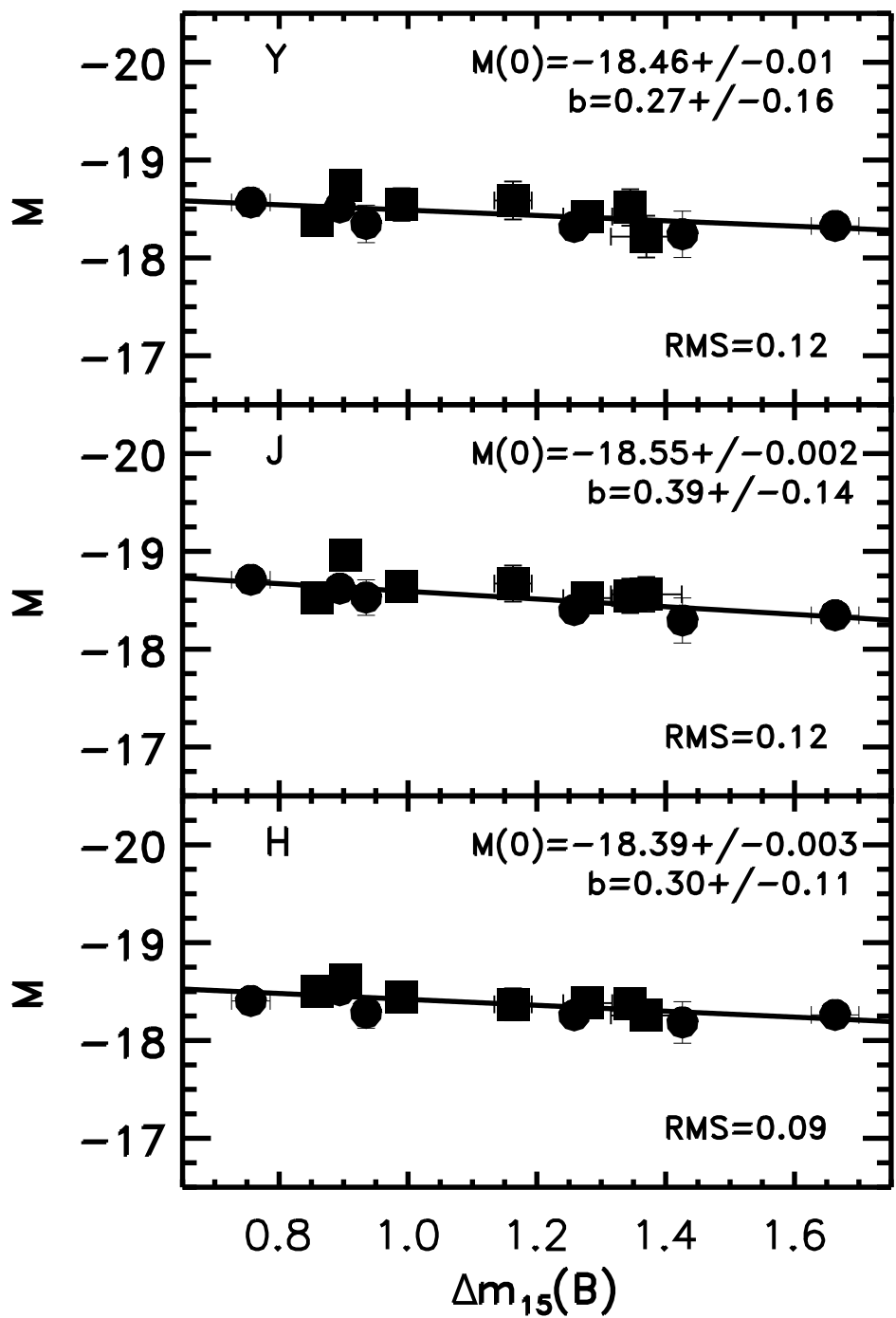}
	}
\end{center}
\caption{Fits of absolute $YJH$ magnitude vs. decline-rate for subsample 3.  The symbols for
  points are the same as those used in Fig. 5.  Uncertainties associated with absolute
  magnitude are smaller than the points, unless shown.  The $Y$-band shows a weak luminosity vs. decline-rate relation, while the $J$- and $H$-bands have a stronger relation, but the dependence of peak magnitude on decline-rate for subsample 3 is significantly less than in optical bands.
\label{fig:fig7} }
\end{figure}

{\it Subsample 3}.  Uses SNe~Ia, with first observations starting within 5 days after NIR peak brightness.  This subsample also omits the highly reddened and fast-declining events.  Like subsample 2, the $Y$-band shows a marginal (1.7$\sigma$) dependence of absolute peak magnitude on decline-rate, while both the $J$- and $H$-bands show a stronger (2.8$\sigma$) correlation between peak absolute magnitude and decline-rate (see Table~\ref{tab:tab5} and Fig. \ref{fig:fig7}). As shown in Figure \ref{fig:fig7} and Table~\ref{tab:tab5}, the rms scatter decreases significantly in $YJH$, compared with subsample 2.

\begin{figure}
\begin{center}
\scalebox{0.75}{
\plotone{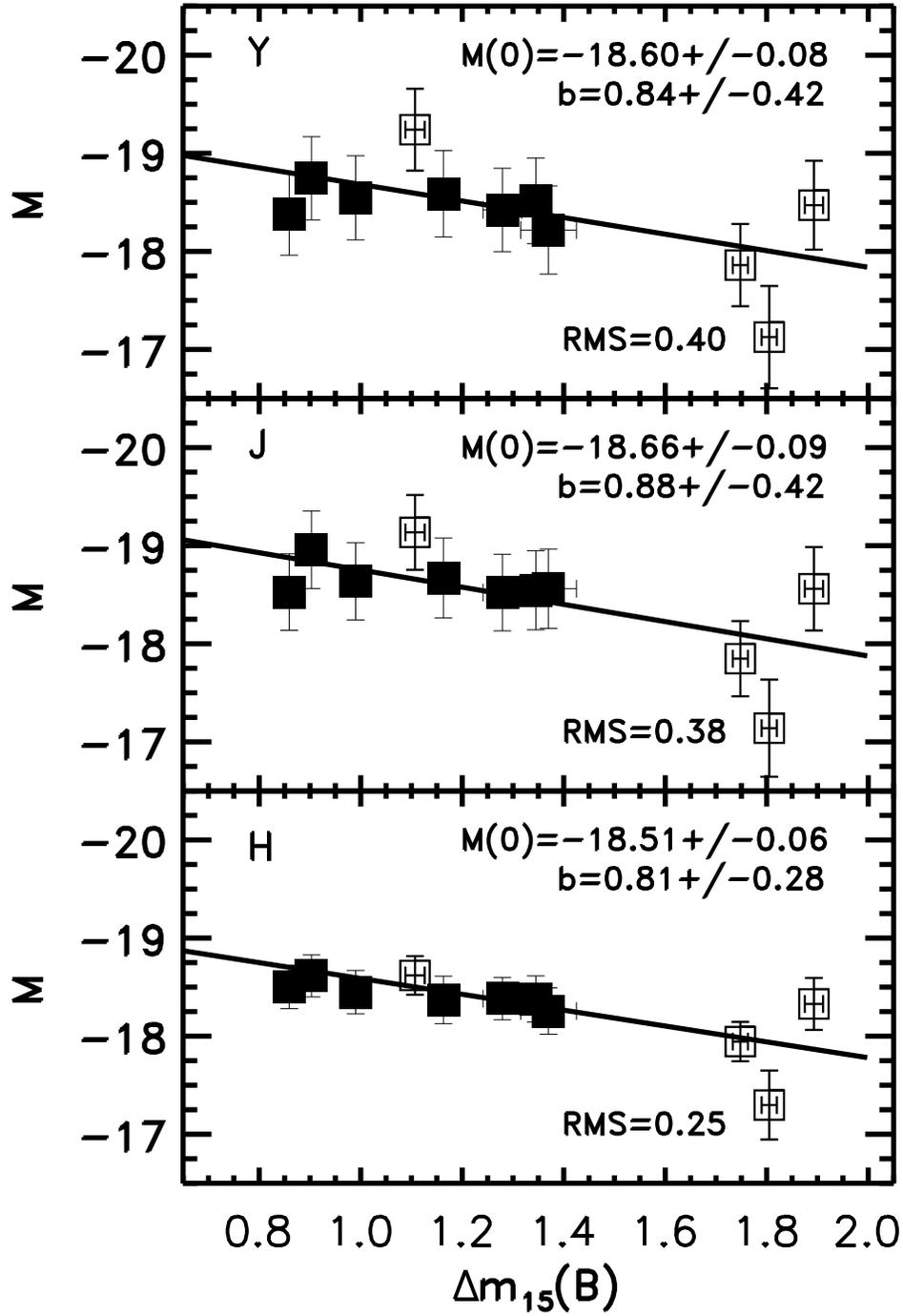}	
	}
\end{center}
\caption{Fits of absolute $YJH$ magnitude vs. decline-rate for subsample 4.  The symbols for points are the same as those used in Fig. 5.  This subsample produces the
largest peak luminosity vs. decline-rate dependence of any set examined.
\label{fig:fig8} }
\end{figure}

{\it Subsample 4}.  Uses the BF subsample, which ensures that purely empirical values for the peak magnitude and $\Delta$$\it{m}$$_{15}$($B$) are used, and avoids the unknown systematics that plague the template fitting process.  This set produces the
largest peak luminosity vs. decline-rate dependence of any set examined (see Fig. \ref{fig:fig8} and Table~\ref{tab:tab5}), but as shown by subsample 5, this is likely due to the inclusion of the highly reddened SN~2006X and the three fast-declining events.

\begin{figure}
\begin{center}
\scalebox{0.75}{
\plotone{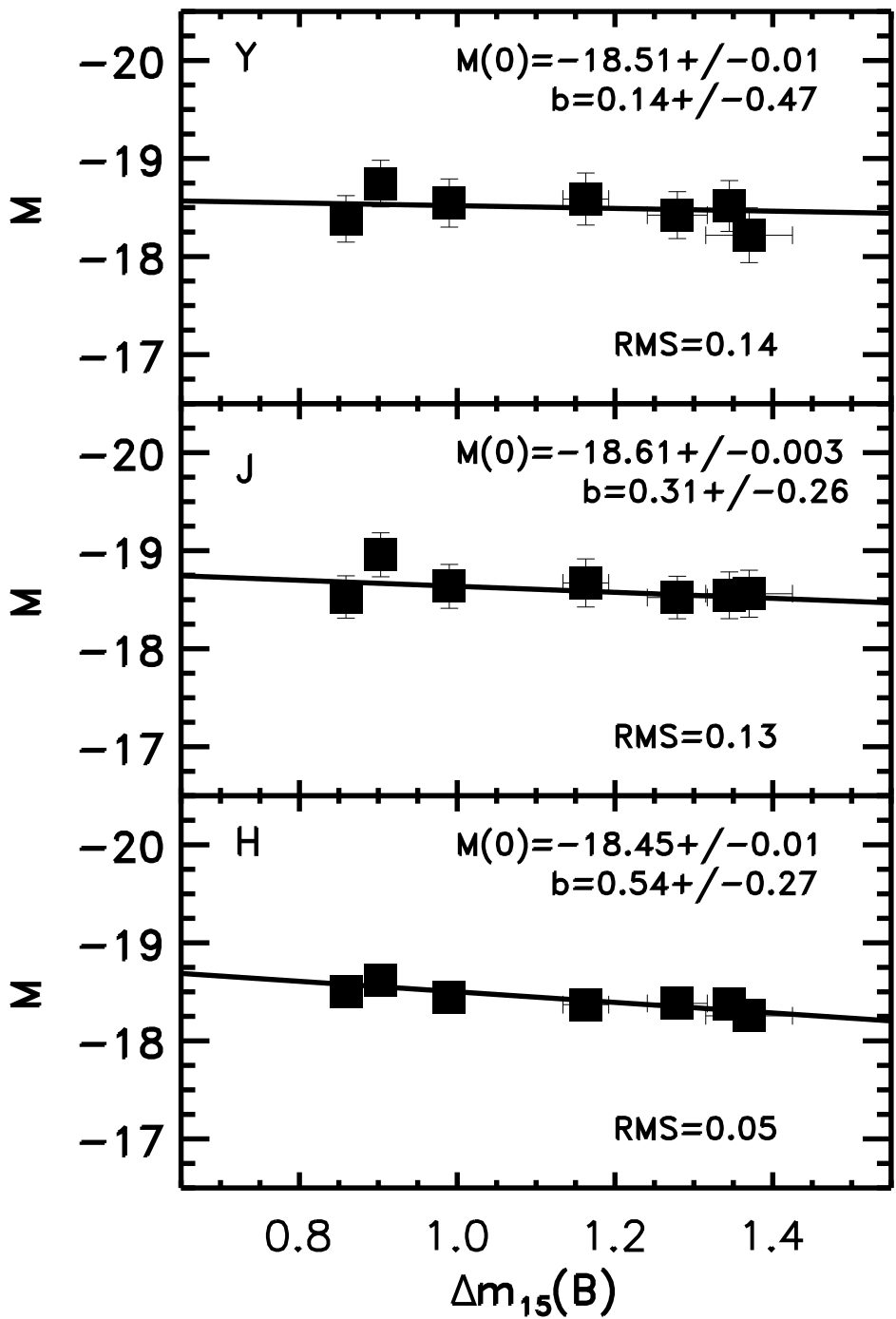}
	}
\end{center}
\caption{Fits of absolute $YJH$ magnitude vs. decline-rate for subsample 5.  The symbols for points are the same as those used in Fig. 5.
  Uncertainties associated with absolute magnitude are smaller than the points, unless
  shown.  There appears to be
little dependence of peak luminosity on decline-rate within the uncertainties
for the $Y$- and $J$-bands and marginal dependence for the $H$-band.  These results are consistent with the luminosity vs. decline-rate relations derived from subsamples 2 and 3.
\label{fig:fig9} }
\end{figure}

{\it Subsample 5}.  Same as subsample 4, but excludes the highly reddened SN~2006X and the fast-declining [$\Delta$$\it{m}$$_{15}$($B$) $>$ 1.7] SNe~Ia.  This subsample has 7 objects with a maximum $\Delta$$\it{m}$$_{15}$($B$) of 1.39.  For this restricted sample, there appears to be only
little dependence of peak luminosity on decline-rate within the uncertainties
for the $Y$- and $J$-bands.  The calculated slopes are 0.14 $\pm$ 0.47 and 0.31 $\pm$ 0.26 for $Y$ and $J$, respectively.
In $H$, there is a marginal ($2\sigma$) detection of a correlation between peak luminosity and decline-rate (see Fig. \ref{fig:fig9}).  These results are consistent with the luminosity vs. decline-rate relations derived from subsamples 2 and 3.  The scatter on the
corrected magnitudes range between 0.05--0.14 mag, with the smallest dispersion
in $H$ and the largest in $Y$.  

\subsection{Comparison with Previous Studies}
To test whether corrections for decline-rate substantially improve the
precision with which SNe~Ia can be used as standardizable candles in the NIR,
one can examine the dispersion in derived absolute magnitude before and after
such corrections are applied.  From examination of Table~\ref{tab:tab6},
which displays both the uncorrected ($b_X = 0$) and corrected rms values given an $\it{R}$$_V$ = 3.1 for the 5 subsamples used throughout the article, we find that a significant rms decrease occurs for the $H$-band, followed by the $J$-band.  The degree of confidence for a non-zero slope in each bandpass [i.e., that a relationship exists between $\Delta$m$_{15}$($B$) and absolute brightness] is indicated in column (6) of Table 6.  Using the t-test statistic, and based on the degrees of freedom for the fit, the confidence level for a non-zero slope was calculated.  For all but subsample 5, we can exclude a zero slope at the 93\% confidence level or better.  This
decrease in the rms and the high confidence levels confirm our earlier findings that correcting observed SNe~Ia values in the $J$- and $H$-bands may substantially improve their utility as cosmological distance indicators. 

To compare with previous studies, we focus on the uncorrected dispersion in Table~\ref{tab:tab6} for subsample 2.  We note that the intrinsic (i.e., uncorrected)
dispersions found for subsample 2 (0.16 mag and 0.17 mag for
$J$ and $H$, respectively) compare well with the rms values of \cite{2009AJ....138.1584K} for SNe~Ia whose NIR light-curves peak before the epoch of $B$$_{max}$ (0.16 mag in $J$ and 0.15 mag in $H$) and of \cite{2008ApJ...689..377W} in $H$ (0.15 mag).  Our $J$-band rms value is much less than the $J$-band rms value of 0.33 mag found by \cite{2008ApJ...689..377W}.\\

\subsection{The Effect of $\it{R}$$_V$ on Derived Correlations}
Given that the value for $\it{R}$$_V$ is a hot topic of debate in SNe~Ia studies, we
examine the effect that lowering $\it{R}$$_V$ (from $\it{R}$$_V$=3.1 to $\it{R}$$_V$=1.7) has on
the slope of the peak absolute $YJH$ magnitudes vs. decline-rate
relation.  The same five subsamples described in $\S$3.2 are used for this investigation.  The measured slopes are given in the
first two rows for each filter and subsample in Table~\ref{tab:tab6}. 

The derived slopes for the two $\it{R}$$_V$ values in subsample 1 are within the measured uncertainties for each NIR band.  Subsample 2
produces the smallest variance between slopes, given different values of
$\it{R}$$_V$.  Subsamples 3 and 4 show a
larger deviation between slopes in all three NIR bands, but all slopes are
within measured uncertainties.
The derived slopes for subsample 5 show a smaller variance than subsamples 1, 3, and 4.  For all five subsamples, the slopes are rather insensitive to the exact value of
$\it{R}$$_V$.

\subsection{Hubble Diagram}
Using the distance modulus derived from equation (1) and the CMB redshifts taken from NASA/IPAC Extragalactic Database (NED), we construct a Hubble diagram, which is plotted in Figure \ref{fig:fig10}.  We use SNe and fit parameters from subsample 2 to derive the distance modulus for each SN in $YJH$ and $B$-bands [M$_B$(0) = $-$19.222(005), b$_B$= 1.042(213)].  The distance moduli for the three NIR bands are averaged for each object.  The resulting Hubble diagram for the combined NIR bands and $B$-band is shown in the top panel of Figure \ref{fig:fig10}.  The solid line represents the adopted concordance model of equation (2).  The residuals between the corrected distance modulus and the standard cosmology are shown in the bottom panels of Figure \ref{fig:fig10}.  We find a combined rms scatter of 0.13 mag, or 6\% in distance, for the NIR bands and a rms scatter of 0.22 mag, or 11\% in distance, for the $B$-band.  The decrease in scatter from optical bands to NIR bands on the Hubble diagram provides strong evidence that SNe~Ia in NIR bands are excellent standard candles.

Note that the dispersion of 0.13 mag obtained in the NIR includes the effects of peculiar velocities.  This was demonstrated by \cite{2010AJ....139..120F}, who pointed out that the residuals in the Hubble diagram in separate filters are highly correlated.  Peculiar velocities also explain why the dispersion in Figure \ref{fig:fig10} decreases with redshift.  Hence, the true precision of SNe~Ia as distance indicators in the NIR is almost certainly better than 6\%.

\section{Bimodal Distribution of Absolute Magnitudes for Fast-Declining SNe~Ia}
Figures \ref{fig:fig5} and \ref{fig:fig8} indicate a range of
peak luminosities for the fast-declining SNe~Ia.  \cite{2009AJ....138.1584K}
suggest that, instead of a range of peak brightnesses, there is a bimodal distribution of peak absolute NIR magnitudes for the fast-declining objects; those with NIR maximum occurring after the time of $B$-band maximum ($\emph{t}$$^{max}_{B}$) are subluminous and those with NIR maximum occurring prior to $\emph{t}$$^{max}_{B}$ have standard brightnesses compared with ``normal" SNe~Ia.  Our data
are consistent with this bimodal distribution hypothesis.  Our sample has three fast-declining SNe~Ia: SN~2005ke
($\Delta$$\it{m}$$_{15}$($B$) = 1.75), SN~2006mr ($\Delta$$\it{m}$$_{15}$($B$)=1.81), and SN~2007on
($\Delta$$\it{m}$$_{15}$($B$)=1.89).  As shown in Table~\ref{tab:tab7}, SNe~2005ke
and 2006mr have an average NIR maxima that occurs $\sim$2-4 days after
$\emph{t}$$^{max}_{B}$ and are, on average, 0.89 mag fainter in $\it{M}$$_Y$, 0.98 mag fainter
in $\it{M}$$_J$, and 0.63 mag fainter in $\it{M}$$_H$ than the average peak luminosity of the
slow to midrange-decliners, respectively, whereas SN~2007on has NIR maximum
that occurs $\sim$2 days prior to $\emph{t}$$^{max}_{B}$ and has standard peak absolute
brightness in all three NIR bands.  It is not surprising that this same bimodality in the absolute magnitude found by \cite{2009AJ....138.1584K} is also found in our sample seeing, as our sample contains some of the same objects that were used in the \cite{2009AJ....138.1584K} study.  It was also found by \cite{2009AJ....138.1584K} that fast-decliners with standard brightnesses, such as SN~2007on, show a distinct secondary maxima in the NIR, much like slow-declining objects, whereas the subluminous fast-decliners have monotonically declining light-curves in all bands.  Our three fast-declining objects show this trend (see Figure \ref{fig:fig2}). 

We re-examine the peak absolute magnitude vs. decline-rate relationship, excluding
the two subluminous fast-decliners and including the standard fast-decliner.
Subsamples 2 and 5 described in $\S$3.2 are used for this analysis, SNe~2005ke and 2006mr are omitted and SN~2007on is included for each subsample.
The fit parameters derived are given in Table~\ref{tab:tab8}.

Including SN~2007on and excluding the two subluminous fast-decliners creates no significant change in slopes for the three NIR bands in subsamples 2 and 5, with the significance on the measured slopes remaining the same. The only case where the fiducial absolute peak magnitude significantly changes when SN~2007on is included is for the $J$-band in subsample 2.  The absolute peak magnitudes for the other bands in subsamples 2 and 5 change very little.  Our findings here confirm the \cite{2009AJ....138.1584K} results that SNe~Ia with NIR maximum occurring prior to $B$-band maximum, such as SN~2007on, have absolute peak magnitudes that are consistent with the absolute peak magnitudes of  slow- to mid-decliners, allowing these SNe~Ia to be used for the peak absolute magnitude vs. decline-rate relationship analysis.

\section{Discussion and Conclusion}
We analyze 27 of the best-observed SNe~Ia from the first 4 years of the
CSP.  Spline curves were fit to the 13 objects with observed maxima in $YJH$.  A
``training set" was created from these 13 SNe~Ia to construct light-curve
templates for the other 14 objects with no observed NIR maxima.  Light-curve
properties obtained from these fits were used to calculate peak absolute
magnitude, and correlations were sought with $B$-band decline-rates.

We discuss difficulties encountered when fitting templates to the NIR light-curves, due to the variations in the strength of the secondary maximum at a given decline-rate.  The problems are especially pronounced
for fast-decliners; however, if the observations begin within 5 days after NIR maximum, SNooPy provides peak magnitudes with an accuracy of $\sim$0.04 mag.  Obtaining additional SNe~Ia light-curves with observed NIR
maxima would be ideal, since splines can be directly fit, but this poses an
observational challenge, since NIR maximum typically occurs several days earlier than it does
in optical bands.  Finding a second light-curve parameter [i.e., other than $\Delta$m$_{15}$(B)]
could certainly be beneficial when fitting NIR light-curve templates to data;
our limited sample size, however, makes such an investigation problematic.
Future larger samples of SNe Ia observed in the NIR could possibly provide
evidence for this second parameter and increase the accuracy of fitting
templates to NIR light-curves, regardless of when the observations begin.

From our original sample of 27 objects, we examine whether a ``Phillips
relation'' exists for the observed data that improves the standardizability of
SNe~Ia in the NIR for five different subsamples.  $\it{R}$$_V$ was kept fixed at 3.1, although the results are shown to not change appreciably if lower values are used.  Consistent with earlier
findings, we find that any peak luminosity vs. decline-rate dependence does indeed
decrease in the NIR compared with the optical, confirming that SNe~Ia are
better ``standard candles" at longer wavelengths.

When fast-decliners and highly reddened events are removed from the sample (e.g., subsample 2), a weaker dependence of absolute peak magnitude on decline-rate is found, with the significance of the measured slope being 1.5$\sigma$ for the $Y$-band and 3$\sigma$ for the $J$- and $H$-bands.  The fact that the rms dispersion decreases significantly when the fast-declining SNe~Ia are removed  confirms the finding by Krisciunas et al. (2009) that a simple linear relationship cannot be used to fit both the slow- to midrange-decliners and the fast-decliners.  For subsample 3, which uses SNe Ia whose first observations begin within 5 days after NIR maxima, we find a weak (1.7$\sigma$) dependence of $Y$ peak luminosity on decline-rate and a stronger (2.8$\sigma$) dependence of $J$- and $H$-peak luminosities on decline-rate.  The dispersion decreases significantly for subsample 3, compared with subsample 2.   
Subsample 4, which uses the BF subsample, shows the largest correlation between peak absolute magnitude and decline-rate, but the fast-declining SNe~Ia strongly affect the slope.  For subsample 5, which uses the BF subsample excluding the fast-decliners and the highly reddened object, there is no significant dependence of peak $Y$ luminosity on decline-rate.  There appears to be a weak (1.2$\sigma$) correlation between decline-rate and the $J$-band absolute peak magnitude in subsample 5 and still a marginal (2.0$\sigma$) correlation between decline-rate and absolute peak magnitude in $H$.  After removal of the highly reddened and fast-declining SNe~Ia, we find evidence at the 2.0-3.0$\sigma$ level for a luminosity vs. decline-rate relations in $J$ and $H$, but minimal evidence for such a relation in $Y$.

 The lowest dispersions in the uncorrected absolute magnitudes are obtained when the highly reddened and fast-declining SNe~Ia are removed from the sample.  Comparing the uncorrected and corrected values, we are at least 93\% confident that the slope is nonzero for subsamples 1-4.  It could be beneficial to correct the peak luminosity in both the $J$- and $H$-bands in order for SNe~Ia to be utilized as precise standard candles in the NIR.  The dispersions obtained are smaller than those found in previous NIR studies.   
 
 As larger samples of SNe~Ia in the NIR are gathered, and especially as more
 objects are observed pre-maximum in the NIR, an interesting study could be to
 investigate whether using a $\Delta$m$_{15}$-like relation, defined in the NIR
 bands (rather than in the $B$-band) would further decrease the dependence of
 absolute peak magnitude on decline-rate.  Our sample presented here does not
 include enough objects that have observed NIR maxima to test this idea, but
 future SNe Ia studies will certainly have more observations that cover the NIR
 peak with greater frequency.

Evidence for a bimodal distribution was found for the NIR absolute peak magnitudes of our
fast-declining SNe~Ia.  The SN from our sample that peaked before the time of $B$-band maximum (SN~2007on) had a ``normal'' peak luminosity, whereas those that
peaked after the time of $B$-band maximum (SNe~2005ke and 2006mr) had subluminous
absolute peak magnitudes compared with the rest of the sample.  It will be interesting to see if future SNe Ia studies find any slow- to midrange-decliners that possess this bimodal distribution.  It would also
be beneficial to obtain larger samples of the fast-declining SNe Ia at NIR, in addition to optical passbands, to further our understanding of these ``peculiar" SNe
Ia.

\begin{figure}
\begin{center}
\scalebox{0.8}{
\plotone{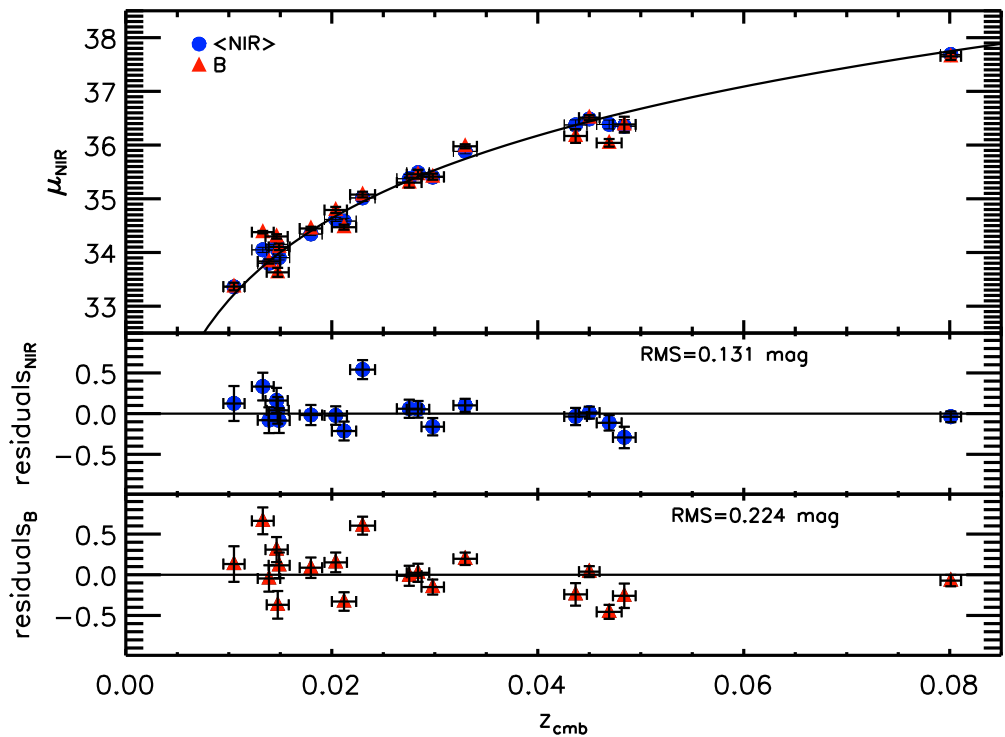}
	}
\end{center}
\caption{\emph{Top}: Hubble diagram constructed from SNe and fit parameters of subsample 2.  The blue circles represent the averaged NIR distance moduli for each SN, and the red triangles represent the $B$-band distace moduli for each SN.  The solid line shows the standard cosmology redshift-distance relationship (eq. [2]).  \emph{Middle}: The NIR residuals with respect to the standard cosmology model.  \emph{Bottom}:  The $B$-band residuals with respect to the standard cosmology model.  
The decrease in scatter from the optical bands to NIR bands provides strong evidence that SNe~Ia in NIR bands are excellent standard candles.
\label{fig:fig10} }
\end{figure}

We also present a Hubble diagram for the NIR bands and the $B$-band, using the parameters derived from subsample 2.  The rms scatter greatly decreases going from optical to NIR bands (0.22 mag in $B$ to 0.13 mag for combined NIR bands), again demonstrating that SNe~Ia observed in NIR bands are excellent standard candles.  The Hubble diagram and Hubble residuals in Figure \ref{fig:fig10}, along with the weaker dependence of absolute peak magnitude on decline-rate found for the NIR bands compared with the optical bands, clearly show the advantages of working in the NIR for distance determinations and cosmological studies.  Future high-$\emph{z}$ SNe Ia cosmological studies may therefore benefit from
observations taken in the rest-frame NIR bands, since SNe~Ia appear to be
better standard candles in these passbands than in the optical (i.e., they
have a weaker dependence of absolute NIR peak magnitude on decline-rate,
systematic uncertainties are minimized due to the reduced effects of dust
extinction, and NIR passbands are relatively insensitive to the exact value of
R$_V$ used).  The $Y$-band, in particular, appears to hold significant promise for
such future studies, since it offers an optimal combination of signal-to-noise ratio,
low sensitivity to dust extinction, and an essentially negligible dependence of
absolute peak magnitude on decline-rate.

\acknowledgements
The authors extend their special thanks to Luis Boldt, Carlos Contreras, Sergio Gonzalez, Wojtek 
Krzeminski, and Francisco Salgado, who worked very hard to gather and reduce the data that
are analyzed in this article.  This work is supported by the National Science Foundation (NSF) 
under grants AST--0306969, AST--0607438, and AST-1008343.  Support for research on the extragalactic distance scale at San Diego State University is provided by proposal number AR$-$10673, provided by NASA through a grant from the Space Telescope Science Institute, which is operated by the Association of Universities for Research in Astronomy, Inc., under NASA contract NAS5$-$26555.  DCL and SK also acknowledge support from NSF grant AST$-$1009571.  M.H. acknowledges support from the Millennium Center for Supernova Science through grant P10$-$064$-$F from Iniciativa Cient\'ifica Milenio, Centro de Astrof\'isica FONDAP 15010003, and Center of Excellence in Astrophysics and Associated Technologies (PFB 06).  G.F.  acknowledges support from FONDECYT grant 3090004.

  \clearpage

\begin{deluxetable}{cccccccl}
\tablecolumns{8}
\tablenum{1} 
\tablewidth{0pt} 
\tablecaption{List of Supernovae and Properties} 
\tablehead{ 
\colhead{SN}& 
\colhead{$\it{z}$$_{helio}$} &
  \colhead{$\it{z}$$_{CMB}$} & 
  \colhead{$\Delta$$\it{m}$$_{15}$($B$)} & 
  \colhead{$\it{E(B-V)}$$_{gal}$} &
  \colhead{$\it{E(B-V)}$$_{host}$} & 
  \colhead{Fit method} & 
  \colhead{Subsample}  \\
  \hline
 \\
  \colhead{(1)}&
  \colhead{(2)}&
  \colhead{(3)}&
  \colhead{(4)}&
  \colhead{(5)}&  
  \colhead{(6)}&  
  \colhead{(7)}&  
  \colhead{(8)}}
  \startdata 
  2004ef & 0.0310 & 0.029 & 1.345(017) & 0.056(006) & 0.097(008) & Template  & 1, 2, 3\\ 
  2004eo & 0.0157
& 0.0147 & 1.370(055) & 0.108(011) & 0.014(018) & Spline & 1, 2, 3, 4, 5\\ 2004ey & 0.0158 &
0.0146 & 0.935(014) & 0.139(014) & $-$0.052(005) & Template & 1, 2, 3\\ 2004gs & 0.0267
& 0.0275 & 1.502(016) & 0.031(003) & 0.157(005) & Template & 1, 2\\ 2004gu & 0.0459
& 0.0469 & 0.900(045) & 0.026(003) & 0.183(009) & Template & 1, 2, 3\\ 2005A & 0.0191
& 0.0184 & 1.224(102) & 0.030(003) & 0.996(016) & Template & 1\\ 2005M & 0.0220
& 0.0230 & 0.859(011) & 0.031(003) & 0.052(011) & Spline & 1, 2, 3, 4, 5\\ 2005ag & 0.0794 &
0.0801 & 0.905(046) & 0.041(004) & 0.047(007) & Template & 1, 2, 3\\ 2005al & 0.0124 &
0.0133 & 1.146(015) & 0.055(005) & $-$0.094(005) & Template & 1, 2, 3\\ 2005am &0.0079 &
0.0090 & 1.509(021) & 0.054(005) & $-$0.030(010) & Template & \\ 2005el & 0.0149
& 0.0149 & 1.345(020) & 0.114(011) & $-$0.043(016) & Spline & 1, 2, 3, 4, 5\\ 2005eq & 0.0290
& 0.0284 & 0.756(022) & 0.074(007) & 0.090(006) & Template & 1, 2, 3\\ 2005hc & 0.0459
& 0.0450 & 0.894(014) & 0.029(003) & 0.048(006) & Template & 1, 2, 3\\ 2005iq & 0.0340
& 0.0330 & 1.258(013) & 0.022(002) & $-$0.040(006) & Template & 1, 2, 3\\ 2005kc &
0.0151 & 0.0139 & 1.163(029) & 0.132(013) & 0.173(006) & Spline & 1, 2, 3, 4, 5\\ 2005ke &
0.0049 & 0.0045 & 1.748(015) & 0.027(003) & 0.036(005) & Spline & 1,  ,      , 4\\ 2005ki &
0.0192 & 0.0204 & 1.279(038) & 0.032(003) & $-$0.018(012) & Spline & 1, 2, 3, 4, 5\\ 2006D &
0.0085 & 0.0097 & 1.388(012) & 0.046(005) & 0.071(010) & Spline & \\ 2006X &
0.0052 & 0.0063 & 1.107(019) & 0.026(003) & 1.202(013) & Spline & 1\\ 2006ax &
0.0167 & 0.0180 & 0.990(016) & 0.050(005) & $-$0.033(006) & Spline & 1, 2, 3, 4, 5\\ 2006bh &
0.0109 & 0.0105 & 1.426(011) & 0.026(003) & $-$0.038(007) & Template & 1, 2, 3\\ 2006eq
& 0.0495 & 0.0484 & 1.660(060) & 0.048(005) & 0.191(033) & Template & 1, 2\\ 2006et
& 0.0222 & 0.0212 & 0.903(012) & 0.019(002) & 0.222(008) & Spline & 1, 2, 3, 4, 5\\ 2006gt &
0.0448 & 0.0437 & 1.663(037) & 0.037(004) & 0.165(019) & Template & 1, 2\\ 2006mr &
0.0059 & 0.0055 & 1.805(015) & 0.021(002) & $-$0.025(013) & Spline & 1, , , 4\\ 2007af &
0.0055 & 0.0063 & 1.203(036) & 0.039(004) & 0.177(019) & Spline & \\ 2007on &
0.0065 & 0.0062 & 1.893(018) & 0.011(001) & 0.022(010) & Spline & 1, , , 4 \\
\enddata
\tablenotetext{\normalsize}{Note. - Col. (1) SN name; Col. (2) Heliocentric
  redshift from the NASA/IPAC Extragalactic Database (NED); Col. (3) Redshift in the
  frame of the 3K CMB (NED); Col. (4) Observed $\Delta$$\it{m}$$_{15}$($B$); Col. (5) Galactic reddening; Col. (6) Host-galaxy reddening; Col. (7)
  Fitting method (spline or template); Col. (8) Subsample(s) SN belongs to (see text and Table 4) If blank, SN is excluded from all subsamples.}
  
\label{tab:tab1}
\end{deluxetable} 
\clearpage

\begin{deluxetable}{ccccc}
\tablecolumns{5} 
\tablenum{2}
\tablewidth{0pt} 
\tablecaption{Derived Distance Moduli} 
\tablehead{ 
\colhead{SN} & 
\colhead{Host-galaxy} &
  \colhead{$\mu$} & 
  \colhead{Distance method} & 
  \colhead{Reference} \\
  \colhead{}&
  \colhead{}&
  \colhead{(mag)}&
  \colhead{}&
  \colhead{}\\
  \hline
 \\
  \colhead{(1)}&
  \colhead{(2)}&
  \colhead{(3)}&
  \colhead{(4)}&
  \colhead{(5)}}
\startdata 
2004ef & UGC 12158 & 35.56 $\pm$ 0.07& LD & \\ 
2004eo & NGC 6928 & 34.00 $\pm$ 0.15 & LD & \\ 
2004ey & UGC 11816 & 33.99 $\pm$ 0.15& LD & \\ 
2004gs & MCG $+$03$-$22$-$020 & 35.31 $\pm$ 0.08 & LD & \\ 
2004gu & FGC 175A & 36.50 $\pm$ 0.05 & LD & \\ 
2005A & NGC 958 & 34.47 $\pm$ 0.12 & LD & \\ 
2005M & NGC 2930 & 34.94 $\pm$ 0.10 & LD & \\ 
2005ag & $\cdots$ & 37.72 $\pm$ 0.03 & LD & \\ 
2005al & NGC 5304 & 33.71 $\pm$ 0.16 & LD & \\ 
2005am & NGC 2811 & 32.85 $\pm$ 0.24 & LD & \\ 
2005el & NGC 1819 & 33.99 $\pm$ 0.14 & LD & \\ 
2005eq & MCG $-$01$-$09$-$006 & 35.43 $\pm$ 0.08 & LD & \\ 
2005hc & MCG $+$00$-$06$-$003 & 36.47 $\pm$ 0.05 & LD & \\ 
2005iq & MCG $-$03$-$01$-$008 & 35.78 $\pm$ 0.07 & LD & \\
2005kc & NGC 7311 & 33.88 $\pm$ 0.16 & LD & \\ 
2005ke & NGC 1371 & 31.84 $\pm$ 0.08 & SBF & \cite{2001ApJ...546..681T}\tablenotemark{*}\\ 
2005ki & NGC 3332 & 34.64 $\pm$ 0.11 & LD & \\ 
2006D & MCG $-$01$-$33$-$34 & 33.00 $\pm$ 0.22 & LD & \\
2006X & NGC 4321 & 30.91 $\pm$ 0.14 & Cepheids & \cite{2001ApJ...553...47F}\\ 
2006ax & NGC 3663 & 34.36 $\pm$ 0.12 & LD & \\ 
2006bh & NGC 7329 & 33.23 $\pm$ 0.21 & LD & \\ 
2006eq & $\cdots$ & 36.64 $\pm$ 0.05 & LD & \\ 
2006et & NGC 232 & 34.80 $\pm$ 0.10 & LD & \\ 
2006gt & 2MASX J00561810$-$0137327 & 36.41 $\pm$ 0.05 & LD & \\ 
2006mr & NGC 1316 & 31.15 $\pm$ 0.23 & SBF & \cite{2001ApJ...559..584A}\\ 
2007af & NGC 5584 & 32.10 $\pm$ 0.34 & LD & \\ 
2007on & NGC 1404 & 31.45 $\pm$ 0.19 & SBF & \cite{2003ApJ...583..712J}\\ 
\enddata
\tablenotetext{\normalsize}{Note. - Col. (1) SN name; Col. (2) Host-galaxy; Col. (3)
  Distance modulus from fits of \textsection{4}; Col. (4) Distance method used (LD -
  luminosity distance, SBF - surface brightness fluctuation); Col. (5) Reference
  used for SBF and Cepheid distances.}
 \tablenotetext{*}{NGC 1371 is a member of the Eridanus group/cluster.  \cite{2001ApJ...546..681T} give a distance modulus of 32.00 $\pm$ 0.08 mag for this group based on SBF measurements of seven members.  Subtracting 0.16 mag (\citealt{2003ApJ...583..712J}) to put this on the \cite{2001ApJ...553...47F} scale gives 31.84 $\pm$ 0.08 mag.}
 
 \label{tab:tab2}
\end{deluxetable} 
\clearpage

\begin{deluxetable}{cccc}
\tablecolumns{4}
\tablenum{3}
\tablewidth{0pt}
\tablecaption{Supernova Light-Curve Properties}
\tablehead{
	\colhead{Filter} &
	\colhead{$\it{t}$$_{max}$ (JD $-$2,453,000)} &
	\colhead{$\it{m}$$_{max}$ (spline)} &
	\colhead{$\it{m}$$_{max}$ (template)}\\
	\colhead{}&
	\colhead{(days)}&
	\colhead{(mag)}&
	\colhead{(mag)}\\
  \hline
 \\
  \colhead{(1)}&
  \colhead{(2)}&
  \colhead{(3)}&
  \colhead{(4)}
}
\startdata
& & \bf{SN~2004ef}\\
$B$ & 264.40 $\pm$ 0.05 & ... & 17.077 $\pm$ 0.007\\
$V$ & 265.88 $\pm$ 0.05 & ... &  16.910 $\pm$ 0.004\\
$Y$ & 261.47 $\pm$ 0.05 & ... & 16.999 $\pm$ 0.008\\
$J$ & 261.44 $\pm$ 0.05 & ... & 17.020 $\pm$ 0.114\\
$H$ & 261.04 $\pm$ 0.05 & ... & 17.409 $\pm$ 0.027\\
 & & \bf{SN~2004eo}\\
$B$ & 278.69 $\pm$ 0.28 & 15.497 $\pm$ 0.013 & 15.552 $\pm$ 0.011\\
$V$ & 280.68 $\pm$ 0.54 & 15.359 $\pm$ 0.012 & 15.346 $\pm$ 0.006\\
$Y$ & 275.85 $\pm$ 0.91 & 15.929 $\pm$ 0.102 & 15.863 $\pm$ 0.027\\
$J$ & 274.85 $\pm$ 0.19 & 15.547 $\pm$ 0.010 & 15.515 $\pm$ 0.025\\
$H$ & 274.18 $\pm$ 0.13 & 15.814 $\pm$ 0.010 & 15.742 $\pm$ 0.024\\
& & \bf{SN~2004ey}\\
$B$ & 304.87 $\pm$ 0.04 & ... & 15.275 $\pm$ 0.004\\
$V$ & 306.20 $\pm$ 0.04 & ... & 15.224 $\pm$ 0.003\\
$Y$ & 300.61 $\pm$ 0.04 & ... & 15.747 $\pm$ 0.033\\
$J$ & 301.40 $\pm$ 0.04 & ... & 15.540 $\pm$ 0.021\\
$H$ & 300.88 $\pm$ 0.04 & ... & 15.757 $\pm$ 0.018\\
& & \bf{SN~2004gs}\\
$B$ & 356.23 $\pm$ 0.07 & ... & 17.277 $\pm$ 0.003\\
$V$ & 357.92 $\pm$ 0.07 & ... & 17.057 $\pm$ 0.004\\
$Y$ & 353.77 $\pm$ 0.07 & ... & 17.241 $\pm$ 0.022\\
$J$ & 353.81 $\pm$ 0.07 & ... & 17.138 $\pm$ 0.032\\
$H$ & 353.52 $\pm$ 0.07 & ... & 17.236 $\pm$ 0.032\\
& & \bf{SN~2004gu}\\
$B$ & 362.56 $\pm$ 0.20 & ... & 17.570 $\pm$ 0.007\\
$V$ & 363.91 $\pm$ 0.20 & ... & 17.401 $\pm$ 0.005\\
$Y$ & 357.98 $\pm$ 0.20 & ... & 17.991 $\pm$ 0.046\\
$J$ & 359.01 $\pm$ 0.20 & ... & 17.934 $\pm$ 0.093\\
$H$ & 358.28 $\pm$ 0.20 & ... & 18.103 $\pm$ 0.050\\
& & \bf{SN~2005A}\\
$B$ & 378.86 $\pm$ 0.24 & ... & 18.239 $\pm$ 0.013\\
$V$ & 380.18 $\pm$ 0.24 & ... & 17.214 $\pm$ 0.010\\
$Y$ & 374.59 $\pm$ 0.24 & ... & 16.306 $\pm$ 0.022\\
$J$ & 375.37 $\pm$ 0.24 & ... & 16.129 $\pm$ 0.023\\
$H$ & 374.85 $\pm$ 0.24 & ... & 16.302 $\pm$ 0.022\\
& & \bf{SN~2005M}\\
$B$ & 406.02 $\pm$ 0.20 & 16.041 $\pm$ 0.004 & 16.048 $\pm$ 0.003\\
$V$ & 407.49 $\pm$ 0.12 & 16.003 $\pm$ 0.010 & 15.998 $\pm$ 0.002\\
$Y$ & 402.52 $\pm$ 0.81 & 16.627 $\pm$ 0.036 & 16.587 $\pm$ 0.006\\
$J$ & 402.59 $\pm$ 0.94 & 16.459 $\pm$ 0.046 & 16.463 $\pm$ 0.013\\
$H$ & 401.38 $\pm$ 0.06 & 16.458 $\pm$ 0.074 & 16.652 $\pm$ 0.010\\
& & \bf{SN~2005ag}\\
$B$ & 414.79 $\pm$ 0.14 & ... & 18.607 $\pm$ 0.005\\
$V$ & 416.22 $\pm$ 0.14 & ... & 18.558 $\pm$ 0.005\\
$Y$ & 410.17 $\pm$ 0.14 & ... & 19.325 $\pm$ 0.032\\
$J$ & 411.10 $\pm$ 0.14 & ... & 19.244 $\pm$ 0.048\\
$H$ & 410.53 $\pm$ 0.14 & ... & 19.044 $\pm$ 0.063\\
& & \bf{SN~2005al}\\
$B$ & 429.86 $\pm$ 0.13 & ... & 15.045 $\pm$ 0.003\\
$V$ & 431.11 $\pm$ 0.13 & ... & 15.094 $\pm$ 0.004\\
$Y$ & 426.42 $\pm$ 0.13 & ... & 15.576 $\pm$ 0.035\\
$J$ & 426.55 $\pm$ 0.13 & ... & 15.355 $\pm$ 0.048\\
$H$ & 426.04 $\pm$ 0.13 & ... & 15.706 $\pm$ 0.021\\
 & & \bf{SN~2005am}\\
 $B$ & 435.90 $\pm$ 0.13 & ... & 13.870 $\pm$ 0.008\\
 $V$ & 437.47 $\pm$ 0.13 & ... & 13.750 $\pm$ 0.006\\
 $Y$ & 433.30 $\pm$ 0.13 & ... & 13.949 $\pm$ 0.032\\
 $J$ & 433.29 $\pm$ 0.13 & ... & 13.746 $\pm$ 0.045\\
 $H$ & 432.97 $\pm$ 0.13 & ... & 14.103 $\pm$ 0.022\\
& & \bf{SN~2005el}\\
$B$ & 647.30 $\pm$ 0.36 & 15.278 $\pm$ 0.011 & 15.307 $\pm$ 0.012\\
$V$ & 647.86 $\pm$ 0.24 & 15.194 $\pm$ 0.012 & 15.261 $\pm$ 0.005\\
$Y$ & 642.66 $\pm$ 0.15 & 15.559 $\pm$ 0.014 & 15.676 $\pm$ 0.012\\
$J$ & 643.15 $\pm$ 0.15 & 15.508 $\pm$ 0.008 & 15.574 $\pm$ 0.015\\
$H$ & 642.33 $\pm$ 0.12 & 15.652 $\pm$ 0.010 & 15.697 $\pm$ 0.012\\
& & \bf{SN~2005eq}\\
$B$ & 654.60 $\pm$ 0.10 & ... & 16.565 $\pm$ 0.004\\
$V$ & 655.93 $\pm$ 0.10 & ... & 16.458 $\pm$ 0.004\\
$Y$ & 650.09 $\pm$ 0.10 & ... & 17.062 $\pm$ 0.020\\
$J$ & 651.11 $\pm$ 0.10 & ... & 16.868 $\pm$ 0.019\\
$H$ & 650.39 $\pm$ 0.10 & ... & 17.122 $\pm$ 0.022\\
& & \bf{SN~2005hc}\\
$B$ & 667.71 $\pm$ 0.07 & ... & 17.431 $\pm$ 0.004\\
$V$ & 669.10 $\pm$ 0.07 & ... & 17.395 $\pm$ 0.004\\
$Y$ & 663.18 $\pm$ 0.07 & ... & 18.049 $\pm$ 0.025\\
$J$ & 664.14 $\pm$ 0.07 & ... & 17.922 $\pm$ 0.026\\
$H$ & 663.55 $\pm$ 0.07 & ... & 18.003 $\pm$ 0.051\\
& & \bf{SN~2005iq}\\
$B$ & 688.03 $\pm$ 0.05 & ... & 16.887 $\pm$ 0.004\\
$V$ & 689.29 $\pm$ 0.05 & ... & 16.902 $\pm$ 0.004\\
$Y$ & 684.42 $\pm$ 0.05 & ... & 17.442 $\pm$ 0.021\\
$J$ & 684.61 $\pm$ 0.05 & ... & 17.365 $\pm$ 0.016\\
$H$ & 684.08 $\pm$ 0.05 & ... & 17.513 $\pm$ 0.025\\
& & \bf{SN~2005kc}\\
$B$ & 698.29 $\pm$ 0.22 & 16.037 $\pm$ 0.004 & 16.063 $\pm$ 0.012\\
$V$ & 699.19 $\pm$ 0.08 & 15.744 $\pm$ 0.004 & 15.745 $\pm$ 0.006\\
$Y$ & 693.97 $\pm$ 0.17 & 15.658 $\pm$ 0.014 & 15.626 $\pm$ 0.014\\
$J$ & 694.69 $\pm$ 0.73 & 15.479 $\pm$ 0.010 & 15.455 $\pm$ 0.016\\
$H$ & 694.10 $\pm$ 0.55 & 15.684 $\pm$ 0.013 & 16.644 $\pm$ 0.010\\
& & \bf{SN~2005ke}\\
$B$ & 698.81 $\pm$ 0.07& 14.882 $\pm$ 0.006 & 14.890 $\pm$ 0.013\\
$V$ & 701.32 $\pm$ 0.24 & 14.189 $\pm$ 0.005 & 14.202 $\pm$ 0.018\\
$Y$ & 701.26 $\pm$ 0.70 & 14.055 $\pm$ 0.018 & 14.078 $\pm$ 0.008\\
$J$ & 700.15 $\pm$ 0.10 & 14.048 $\pm$ 0.006 & 14.016 $\pm$ 0.010\\
$H$ & 702.07 $\pm$ 0.13 & 13.932 $\pm$ 0.008 & 13.928 $\pm$ 0.004\\
& & \bf{SN~2005ki}\\
$B$ & 706.13 $\pm$ 1.28 & 15.689 $\pm$ 0.011 & 15.690 $\pm$ 0.009\\
$V$ & 706.40 $\pm$ 0.19 & 15.670 $\pm$ 0.005 & 15.698 $\pm$ 0.007\\
$Y$ & 701.62 $\pm$ 0.06 & 16.233 $\pm$ 0.011 & 16.278 $\pm$ 0.012\\
$J$ & 702.73 $\pm$ 0.26 & 16.129 $\pm$ 0.012 & 16.186 $\pm$ 0.020\\
$H$ & 701.48 $\pm$ 0.75 & 16.265 $\pm$ 0.028 & 16.321 $\pm$ 0.024\\
& & \bf{SN~2006D}\\
$B$ & 757.75 $\pm$ 0.06 & 14.319 $\pm$ 0.007 & 14.339 $\pm$ 0.008\\
$V$ & 759.13 $\pm$ 0.19 & 14.183 $\pm$ 0.007 & 14.212 $\pm$ 0.004\\
$Y$ & 754.85 $\pm$ 0.05& 14.754 $\pm$ 0.117 & 14.630 $\pm$ 0.013\\
$J$ & 755.05 $\pm$ 0.067 & 14.379 $\pm$ 0.008 & 14.460 $\pm$ 0.010\\
$H$ & 754.90 $\pm$ 0.055 & 14.500 $\pm$ 0.007 & 14.590 $\pm$ 0.008\\
& & \bf{SN~2006X}\\
$B$ & 786.23 $\pm$ 0.14 & 15.324 $\pm$ 0.007 & 15.383 $\pm$ 0.021\\
$V$ & 789.18 $\pm$ 0.13 & 14.111 $\pm$ 0.011 & 14.111 $\pm$ 0.010\\
$Y$ & 781.50 $\pm$ 0.11 & 13.144 $\pm$ 0.007 & 13.175 $\pm$ 0.017\\
$J$ & 782.74 $\pm$ 0.12 & 12.866 $\pm$ 0.006 & 12.901 $\pm$ 0.012\\
$H$ & 781.76 $\pm$ 0.12 & 12.996 $\pm$ 0.004 & 12.999 $\pm$0.010\\
& & \bf{SN~2006ax}\\
$B$ & 827.59 $\pm$ 0.13 & 15.201 $\pm$ 0.005 & 15.213 $\pm$ 0.007\\
$V$ & 828.27 $\pm$ 0.36 & 15.211 $\pm$ 0.002 & 15.228 $\pm$ 0.005\\
$Y$ & 823.47 $\pm$ 0.57 & 15.835 $\pm$ 0.012 & 15.932 $\pm$ 0.014\\
$J$ & 823.72 $\pm$ 0.55 & 15.740 $\pm$ 0.011 & 15.754 $\pm$ 0.012\\
$H$ & 824.24 $\pm$ 0.40 & 15.922 $\pm$ 0.021 & 16.046 $\pm$ 0.010\\
& & \bf{SN~2006bh}\\
$B$ & 833.65 $\pm$ 0.04 & ... & 14.488 $\pm$ 0.006\\
$V$ & 835.00 $\pm$ 0.04 & ... & 14.477 $\pm$ 0.004\\
$Y$ & 830.57 $\pm$ 0.04 & ... & 14.978 $\pm$ 0.009\\
$J$ & 830.57 $\pm$ 0.04 & ... & 14.930 $\pm$ 0.018\\
$H$ & 830.13 $\pm$ 0.04 & ... & 15.043 $\pm$ 0.014\\
 & & \bf{SN~2006eq}\\
 $B$ & 976.72 $\pm$ 0.43 & ... & 18.726 $\pm$ 0.027\\
 $V$ & 978.73 $\pm$ 0.43 & ... & 18.438 $\pm$ 0.019\\
 $Y$ & 975.28 $\pm$ 0.43 & ... & 18.236 $\pm$ 0.018\\
 $J$ & 975.60 $\pm$ 0.43 & ... & 18.247 $\pm$ 0.114\\
 $H$ & 975.55 $\pm$ 0.43 & ... & 18.374 $\pm$ 0.065\\
& & \bf{SN~2006et}\\
$B$ & 993.55 $\pm$ 0.37 & 16.038 $\pm$ 0.004 & 16.060 $\pm$ 0.007\\
$V$ & 995.75 $\pm$ 0.21 & 15.859 $\pm$ 0.008 & 15.865 $\pm$ 0.004\\
$Y$ &  989.27 $\pm$ 1.07 & 16.344 $\pm$ 0.016 & 16.333 $\pm$ 0.013\\
$J$ & 990.48 $\pm$ 0.53 & 16.055 $\pm$ 0.063 & 16.113 $\pm$ 0.012\\
$H$ & 989.34 $\pm$ 0.27 & 16.323 $\pm$ 0.031 & 16.401 $\pm$ 0.019\\
& & \bf{SN~2006gt}\\
$B$ & 1003.15 $\pm$ 0.10 & ... & 18.383 $\pm$ 0.016\\
$V$ & 1005.19 $\pm$ 0.10 & ... & 18.129 $\pm$ 0.011\\
$Y$ & 1001.95 $\pm$ 0.10 & ... & 18.326 $\pm$ 0.015\\
$J$ & 1002.28 $\pm$ 0.10 & ... & 18.244 $\pm$ 0.034\\
$H$ & 1002.29 $\pm$ 0.10 & ... & 18.265 $\pm$ 0.035\\
& & \bf{SN~2006mr}\\
$B$ & 1050.66 $\pm$ 0.08 & 15.387 $\pm$ 0.009 & 15.555 $\pm$ 0.019\\
$V$ & 1052.53 $\pm$ 0.04 & 14.640 $\pm$ 0.009 & 14.780 $\pm$ 0.016\\
$Y$ & 1056.68 $\pm$ 0.06 & 14.019 $\pm$ 0.006 & 14.075 $\pm$ 0.008\\
$J$ & 1054.31 $\pm$ 0.02 & 14.007 $\pm$ 0.004 & 14.107 $\pm$ 0.014\\
$H$ & 1055.30 $\pm$ 0.03 & 13.850 $\pm$ 0.005 & 13.903 $\pm$ 0.009\\
& & \bf{SN~2007af}\\
$B$ & 1174.78 $\pm$ 0.20 & 13.440 $\pm$ 0.011 & 13.450 $\pm$ 0.012\\
$V$ & 1176.16 $\pm$ 0.17 & 13.228 $\pm$ 0.016 & 13.235 $\pm$ 0.007\\
$Y$ & 1170.57 $\pm$ 0.08 & 13.582 $\pm$ 0.009 & 13.584 $\pm$ 0.007\\
$J$ & 1171.12 $\pm$ 0.04 & 13.469 $\pm$ 0.002 & 13.460 $\pm$ 0.008\\
$H$ & 1170.21 $\pm$ 0.15 & 13.623 $\pm$ 0.004 & 13.625 $\pm$ 0.005\\
& & \bf{SN~2007on}\\
$B$ & 1420.50 $\pm$ 0.08 & 13.085 $\pm$ 0.007 & 13.187 $\pm$ 0.015\\
$V$ & 1421.78 $\pm$ 0.15 & 12.973 $\pm$ 0.007 & 13.024 $\pm$ 0.009\\
$Y$ & 1418.04 $\pm$ 0.04 & 13.254 $\pm$ 0.005 & 13.388 $\pm$ 0.010\\
$J$ & 1418.68 $\pm$ 0.32 & 13.093 $\pm$ 0.063 & 13.239 $\pm$ 0.016\\
$H$ & 1418.36 $\pm$ 0.08 & 13.253 $\pm$ 0.006 & 13.285 $\pm$ 0.007\\
\enddata
\tablenotetext{\normalsize}{Note. - Col. (1) Filter; Col. (2) $\it{t}$$_{max}$; Col. (3) $\emph{K}$-corrected $\it{m}$$_{max}$ spline; Col. (4) $\emph{K}$-corrected $\it{m}$$_{max}$ template.}
\label{tab:tab3}
\end{deluxetable}
\clearpage

\begin{deluxetable}{ccc}
\tablecolumns{3}
\tablenum{4}
\tablewidth{0pt}
\tablecaption{Description of Subsamples}
\tablehead{
	\colhead{Subsample}&
	\colhead{Description}&
	\colhead{$\it{N}$$_{SNe}$}
}
\startdata
1 & All 24 SNe & 24\\
 2 & Exc. fast-declining ($\Delta$$\it{m}$$_{15}$($B$) $>$ 1.7) and highly reddened ($\it{E(B-V)}$ $>$ 0.9) & 19 \\
3 & SNe whose first observation begins $\leq$ 5 days after NIR maximum \\
& exc. fast-declining and highly reddened events & 13\\
4 & Best fit & 11 \\
5 & Best fit exc. fast-declining and highly reddened & 7
\enddata
\label{tab:tab4}
\end{deluxetable}
\clearpage

\begin{deluxetable}{cccccc}   
\tablecolumns{6} 
\tablenum{5} 
\tablewidth{0pt} 
\tabletypesize{\small} 
\tablecaption{Fit Parameters of Peak NIR Magnitude Versus $\Delta$$\it{m}$$_{15}$($B$)} 
\tablehead{
  \colhead{Filter $\it{X}$} & 
  \colhead{$\it{M}$$_{X}$(0)} & 
  \colhead{$\it{b}$$_{X}$} &
 \colhead{$\it{RMS}$} & 
 \colhead{$\it{N}$$_{SNe}$} & 
 \colhead{Subsample} \\
 \colhead{}&
 \colhead{(mag)}&
 \colhead{(mag decline-rate$^{-1}$)}&
 \colhead{(mag)}&
 \colhead{}&
 \colhead{}\\
  \hline
 \\
  \colhead{(1)}&
  \colhead{(2)}&
  \colhead{(3)}&
  \colhead{(4)}&
  \colhead{(5)}&  
  \colhead{(6)}
 }
   \startdata  
$Y$ & $-$18.535 $\pm$ 0.032 & 0.611 $\pm$ 0.263 & 0.382 & 24 &1\\ 
$J$ & $-$18.588 $\pm$ 0.028 & 0.691 $\pm$ 0.230 & 0.336 & 24 & 1\\
$H$ & $-$18.432 $\pm$ 0.017 & 0.581 $\pm$ 0.153 & 0.235 & 24 & 1\\

$Y$ & $-$18.494 $\pm$ 0.007 & 0.294 $\pm$ 0.187 & 0.184 & 19 & 2\\ 
$J$ & $-$18.574 $\pm$ 0.003 & 0.363 $\pm$ 0.119 & 0.141 & 19 & 2\\
 $H$ & $-$18.415 $\pm$ 0.006 & 0.459 $\pm$ 0.140 & 0.142 & 19 & 2\\

$Y$ & $-$18.461 $\pm$ 0.005 & 0.273 $\pm$ 0.160 & 0.122 & 13 & 3\\ 
$J$ & $-$18.552 $\pm$ 0.002 & 0.392 $\pm$ 0.141 & 0.123 & 13 & 3\\ 
$H$ & $-$18.390 $\pm$ 0.003 & 0.302 $\pm$ 0.109 & 0.086 & 13 & 3\\

$Y$ & $-$18.598 $\pm$ 0.084 & 0.844 $\pm$ 0.418 & 0.396 & 11 & 4\\ 
$J$ & $-$18.665 $\pm$ 0.092 & 0.878 $\pm$ 0.417 & 0.384 & 11 & 4\\ 
$H$ & $-$18.507 $\pm$ 0.056 & 0.808 $\pm$ 0.276 & 0.250 & 11 & 4\\

$Y$ & $-$18.505 $\pm$ 0.007 & 0.139 $\pm$ 0.471 & 0.136 & 7 & 5\\ 
$J$ & $-$18.606 $\pm$ 0.003 & 0.307 $\pm$ 0.261 & 0.125 & 7 & 5\\ 
$H$ & $-$18.446 $\pm$ 0.009 & 0.538 $\pm$ 0.273 & 0.053 & 7 & 5\\

\enddata
\tablenotetext{\normalsize}{Note. - Col. (1) Filter;
  Col. (2) Absolute magnitude for $\Delta$$\it{m}$$_{15}$($B$) = 1.1 and no reddening; Col. (3)
  Luminosity vs. decline-rate slope;
  Col. (4) rms of fit in magnitudes; Col. (5) Number of SNe used in fits; Col. (6) Sample of
  SNe used in fit (see text).}
  \label{tab:tab5}
\end{deluxetable} 
\clearpage

\begin{deluxetable}{ccccccc}   
\tablecolumns{7} 
\tablenum{6} 
\tablewidth{0pt} 
\tabletypesize{\small} 
\tablecaption{Fit Parameters}
\tablehead{
 \colhead{Filter}& 
 \colhead{$\it{M}$$_X$(0)}& 
 \colhead{$\it{b}$$_X$}&
  \colhead{$\it{R}$$_V$}& 
  \colhead{$\it{RMS}$}& 
  \colhead{Confidence level}& 
  \colhead{Subsample}\\
  \colhead{}&
  \colhead{(mag)}&
  \colhead{(mag decline-rate$^{-1}$)}&
  \colhead{}&
  \colhead{(mag)}&
  \colhead{(\%)}&
  \colhead{}\\
  \hline
 \\
  \colhead{(1)}&
  \colhead{(2)}&
  \colhead{(3)}&
  \colhead{(4)}&
  \colhead{(5)}&  
  \colhead{(6)}&  
  \colhead{(7)}
}  
\startdata 
$Y$ & $-$18.418 $\pm$ 0.019 & 0.495 $\pm$ 0.172 & 1.7 & 0.262 & 99.6 & 1\\
$Y$ & $-$18.535 $\pm$ 0.032 & 0.611 $\pm$ 0.263 & 3.1 & 0.382 & 98.4 & 1\\ 
$Y$ & $-$18.530 $\pm$ 0.000 & 0.000 & 3.1 & 0.424 & $-$ & 1\\
& & & & &\\
$J$ & $-$18.490 $\pm$ 0.020 & 0.606 $\pm$ 0.170 & 1.7 & 0.258 & 99.9 & 1\\
$J$ & $-$18.588 $\pm$ 0.028 & 0.691 $\pm$ 0.230 & 3.1 & 0.336 & 99.7 & 1\\
$J$ & $-$18.560 $\pm$ 0.000 & 0.000 & 3.1 & 0.399 & $-$ & 1\\
& & & & &\\
$H$ & $-$18.366 $\pm$ 0.014 & 0.545 $\pm$ 0.131 & 1.7 & 0.201 & 99.9 & 1\\
$H$ & $-$18.432 $\pm$ 0.017 & 0.581 $\pm$ 0.153 & 3.1 & 0.235 & 99.9 & 1\\
$H$ & $-$18.389 $\pm$ 0.000 & 0.000 & 3.1 & 0.292 & $-$ & 1\\
& & & & &\\ 
$Y$ & $-$18.441 $\pm$ 0.005 & 0.413 $\pm$ 0.146 & 1.7 & 0.145 & 99.4 & 2\\ 
$Y$ & $-$18.494 $\pm$ 0.007 & 0.294 $\pm$ 0.187 & 3.1 & 0.184 & 93.6 & 2\\ 
$Y$ & $-$18.503 $\pm$ 0.000 & 0.000 & 3.1 & 0.189 & $-$ & 2\\ 
& & & & &\\
 $J$ & $-$18.514 $\pm$ 0.002 & 0.376 $\pm$ 0.093 & 1.7 & 0.116 & 99.9 & 2\\ 
 $J$ & $-$18.574 $\pm$ 0.003 & 0.363 $\pm$ 0.119 & 3.1 & 0.141 & 99.6 & 2\\ 
 $J$ & $-$18.545 $\pm$ 0.000 & 0.000 & 3.1 & 0.163 & $-$ & 2\\ 
 & & & & &\\ 
 $H$ & $-$18.390 $\pm$ 0.006 & 0.469 $\pm$ 0.130 & 1.7 & 0.128 & 99.9 & 2\\ 
 $H$ & $-$18.415 $\pm$ 0.006 & 0.459 $\pm$ 0.140 & 3.1 & 0.142 & 99.8 & 2\\ 
 $H$ & $-$18.398 $\pm$ 0.000 & 0.000 & 3.1 & 0.169 & $-$ & 2\\ 
 & & & & &\\
 $Y$ & $-$18.419 $\pm$ 0.005 & 0.306 $\pm$ 0.115 & 1.7 & 0.098 & 99.0 & 3\\
 $Y$ & $-$18.461 $\pm$ 0.005 & 0.273 $\pm$ 0.160 & 3.1 & 0.122 & 94.1 & 3\\
 $Y$ & $-$18.443 $\pm$ 0.000 & 0.000 & 3.1 & 0.147 & $-$ & 3\\
 & & & & &\\
 $J$ & $-$18.525 $\pm$ 0.003 & 0.394 $\pm$ 0.113 & 1.7 & 0.101 & 99.7 & 3\\
 $J$ & $-$18.552 $\pm$ 0.002 & 0.392 $\pm$ 0.141 & 3.1 & 0.123 & 99.0 & 3\\
 $J$ & $-$18.544 $\pm$ 0.000 & 0.000 & 3.1 & 0.164 & $-$ & 3\\
 & & & & &\\
 $H$ & $-$18.371 $\pm$ 0.004 & 0.309 $\pm$ 0.097 & 1.7 & 0.073 & 99.6 & 3\\
 $H$ & $-$18.390 $\pm$ 0.003 & 0.302 $\pm$ 0.109 & 3.1& 0.086 & 99.1 & 3\\
 $H$ & $-$18.375 $\pm$ 0.000 & 0.000 & 3.1 & 0.119 & $-$ & 3\\
 & & & & &\\
 $Y$ & $-$18.404 $\pm$ 0.063 & 0.605 $\pm$ 0.323 & 1.7 & 0.302 & 95.5 & 4\\
 $Y$ & $-$18.598 $\pm$ 0.084 & 0.844 $\pm$ 0.418 & 3.1 & 0.396 & 96.2 & 4\\
 $Y$ & $-$18.530 $\pm$ 0.000 & 0.000 & 3.1 & 0.508 & $-$ & 4\\
 & & & & &\\
 $J$ & $-$18.494 $\pm$ 0.075 & 0.746 $\pm$ 0.353 & 1.7 & 0.222 & 96.7 & 4\\
 $J$ & $-$18.665 $\pm$ 0.092 & 0.878 $\pm$ 0.417 & 3.1 & 0.384 & 96.7 & 4\\ 
 $J$ & $-$18.571 $\pm$ 0.000 & 0.000 & 3.1 & 0.516 & $-$ & 4\\
 & & & & &\\
 $H$ & $-$18.339 $\pm$ 0.050 & 0.591 $\pm$ 0.254 & 1.7 & 0.236 & 97.6 & 4\\
 $H$ & $-$18.507 $\pm$ 0.056 & 0.808 $\pm$ 0.276 & 3.1 & 0.250 & 99.1 & 4\\
 $H$ & $-$18.363 $\pm$ 0.000 & 0.000 & 3.1 & 0.359 & $-$ & 4\\ 
 & & & & &\\ 
 $Y$ & $-$18.464 $\pm$ 0.006 & 0.004 $\pm$ 0.425 & 1.7 & 0.120 & $-$ & 5\\ 
 $Y$ & $-$18.505 $\pm$ 0.007 & 0.139 $\pm$ 0.471 & 3.1 & 0.136 & $-$ & 5\\ 
$Y$ & $-$18.507 $\pm$ 0.000 & 0.000 & 3.1 & 0.155 & $-$ & 5\\ 
& & & & &\\ 
$J$ & $-$18.591 $\pm$ 0.003 & 0.245 $\pm$ 0.230 & 1.7 & 0.101 & 83.9 & 5\\ 
$J$ & $-$18.606 $\pm$ 0.005 & 0.307 $\pm$ 0.261 & 3.1 & 0.125 & 85.8 & 5\\ 
$J$ & $-$18.628 $\pm$ 0.000 & 0.000 & 3.1 & 0.143 & $-$ & 5\\ 
& & & & &\\ 
$H$ & $-$18.428 $\pm$ 0.008 & 0.444 $\pm$ 0.270 & 1.7 & 0.054 & 91.5 & 5\\ 
$H$ & $-$18.446 $\pm$ 0.009 & 0.538 $\pm$ 0.273 & 3.1 & 0.053 & 94.9 & 5\\ 
$H$ & $-$18.442 $\pm$ 0.000 & 0.000 & 3.1 & 0.106 & $-$ & 5\\
 \enddata
\tablenotetext{\normalsize}{Note. - Col. (1) Filter; Col. (2) Absolute magnitude
  for $\Delta$$\it{m}$$_{15}$($B$) = 1.1 and no reddening; Col. (3) Luminosity vs. decline-rate
  slope; Col. (4) Total- to-selective absorption coefficient; Col. (5) rms of fit in
  magnitudes; Col. (6) Percent confidence level that slope is not zero; Col. (7) Sample.}
  \label{tab:tab6}
\end{deluxetable}
\clearpage

\begin{deluxetable}{ccccc}   
\tablecolumns{5}
\tablenum{7}
 \tablewidth{0pt} 
\tablecaption{NIR Time of Maximum and Absolute Magnitude at Maximum}
\tablehead{ 
	 \colhead{Filter}& 
	 \colhead{$\emph{t}$$^{max}_{B}$}&
  	\colhead{$\langle$$\it{M}$$\rangle$}& 
	\colhead{$\sigma$}& 
	\colhead{$\it{N}$$_{SNe}$}\\
	\colhead{}&
	\colhead{(days)}&
	\colhead{(mag)}&
	\colhead{(mag)}&
	\colhead{}\\
  \hline
 \\
  \colhead{(1)}&
  \colhead{(2)}&
  \colhead{(3)}&
  \colhead{(4)}&
  \colhead{(5)}
    
} 
\startdata
& & Slow to mid-decliners & &\\
$Y$ & $-$3.649 & $-$18.564 & 0.129 & 21\\
$J$ & $-$3.178 & $-$18.651 & 0.135 & 21\\
$H$ & $-$3.735 & $-$18.425 & 0.130 & 21\\
& & Early-Peaking Fast-Decliners & &\\
$Y$ & $-$2.500 & $-$18.471 & 0.190 & 1\\
$J$ & $-$1.820 & $-$18.561 & 0.200 & 1\\
$H$ & $-$2.140 & $-$18.328 & 0.190 & 1\\
& & Late-Peaking Fast-Decliners & &\\
$Y$ & 4.230 &  $-$17.668 & 0.127 & 2\\
$J$ & 2.490 & $-$17.669 & 0.125 & 2\\
$H$ & 3.950 & $-$17.797 & 0.125 & 2\\
\enddata
\tablenotetext{\normalsize}{Note. - Col. (1) Filter;
  Col. (2) Time of NIR maximum; Col. (3) Average absolute peak magnitude; Col. (4) rms of fit in magnitudes;
  Col. (5) Number of SNe used in fits.}
  \label{tab:tab7}
\end{deluxetable}
\clearpage

\begin{deluxetable}{ccccccc}   
\tablecolumns{7} 
\tablenum{8} 
\tablewidth{0pt} 
\tabletypesize{\small} 
\tablecaption{Fit Parameters of Peak Magnitude Versus $\Delta$$\it{m}$$_{15}$($B$) for a Bimodal Distribution} 
\tablehead{
  \colhead{Filter $\it{X}$} & 
  \colhead{$\it{M}$$_{X}$(0)} & 
  \colhead{$\it{b}$$_{X}$} & 
  \colhead{$\it{RMS}$} & 
  \colhead{$\it{N}$$_{SNe}$} & 
  \colhead{Subsample} \\
  \colhead{}&
  \colhead{(mag)}&
  \colhead{(mag decline-rate$^{-1}$)}&
  \colhead{(mag)}&
  \colhead{}&
  \colhead{}\\
  \hline
 \\
  \colhead{(1)}&
  \colhead{(2)}&
  \colhead{(3)}&
  \colhead{(4)}&
  \colhead{(5)}&  
  \colhead{(6)}
  }
   \startdata 
$Y$ & $-$18.497 $\pm$ 0.010 & 0.270 $\pm$ 0.160 & 0.181 & 20 & 2\\ 
$J$ & $-$18.551 $\pm$ 0.005 & 0.332 $\pm$ 0.112 & 0.146 & 20 & 2\\ 
$H$ & $-$18.418 $\pm$ 0.008 & 0.417 $\pm$ 0.123 & 0.145 & 20 & 2\\

$Y$ & $-$18.507 $\pm$ 0.022 & 0.098 $\pm$ 0.261 & 0.137 & 8 & 5\\ 
$J$ & $-$18.617 $\pm$ 0.013 & 0.240 $\pm$ 0.178 & 0.124 & 8 & 5\\ 
$H$ & $-$18.451 $\pm$ 0.012 & 0.330 $\pm$ 0.173 & 0.074 & 8 & 5\\

\enddata
\tablenotetext{\normalsize}{Note. - Col. (1) Filter;
  Col. (2) Absolute magnitude for $\Delta$$\it{m}$$_{15}$($B$) = 1.1 and no reddening; Col. (3)
  Luminosity vs. decline-rate slope; Col. (4) rms of fit in magnitudes; Col. (5) Number of SNe used in fits; Col. (6) Sample of SNe used in fit (see text).}
  \label{tab:tab8}
\end{deluxetable} 
\clearpage

\clearpage

\end{document}